\documentclass[12pt] {iopart}
\usepackage{amsmath}
\usepackage{cite}
\usepackage{xcolor}
\usepackage{graphicx}
\begin{document}
\title[Exploring Shell Evolution and N = 40 Magicity]{Exploring Shell Evolution and N = 40 Magicity in Light-Mass Nuclei with Relativistc Mean Field Approach}

\author{Priyanka$^1$, Praveen K. Yadav$^2$, M. S. Mehta$^1$ and M. Bhuyan$^3$}

\address{$^1$ Department of Physics, Rayat Bahra University, Mohali 140104, India}
\address{$^2$ Department of Physics and Material Science, Thapar Institute of Engineering 
and Technology, Patiala-147004, Punjab, India}
\address{$^3$ Institute of Physics, Sachivalaya Marg, Bhubaneswar 751005, Odisha, India}

\ead{priya80593@gmail.com, praveenkumarneer@gmail.com, mehta\_iop@yahoo.co.uk, mrutunjaya.b@iopb.res.in}

\begin{abstract}
We employ the relativistic mean-field (RMF) approach with NL3 
parameters to study shell and sub-shell closures in the 
isotopic chains of Cl, Ar, K, Ca, Sc, Ti, V, and Cr nuclei. By 
analyzing nuclear bulk properties—binding energy, charge 
radii, two-neutron separation energies, deformation parameters 
($\beta_2$), and single-particle levels—we trace the evolution 
of magic numbers. Our results highlight the single-particle 
energy levels to examine nuclear shell closure and the 
occupancy of individual nucleon orbitals. A comprehensive 
picture of N = 34 shell closure is particularly prominent in 
the isotopic chains of Cl, Ar, and Ti nuclei, where the 
magicity associated with this number remains evident across 
several isotopes. In contrast, the N = 40 exhibits a more 
robust and widespread manifestation across all the examined 
nuclei, indicating that its shell closure is less sensitive 
to the specific isotopic environment and more universally 
applicable across the given nuclear systems. To further 
validate these closures, we apply the coherent density 
fluctuation model (CDFM) to assess isospin-dependent 
observables, such as the symmetry energy and its surface and 
volume components. A systematic analysis of the symmetry 
energy, computed using the relativistic mean-field (RMF) 
approach, maps the evolution of the shell structure at 
$N = 20$ and 28, supports sub-magicity at $N = 34$, and 
strongly suggests a shell closure at $N = 40$, reflected 
consistently in both bulk and surface properties. The results 
of bulk properties are compared with available experimental 
data, and we find good agreement with observed trends in the 
nuclear structure of these isotopes. The interplay between 
shell structure and nuclear deformation remains a central 
topic of research, and future theoretical and experimental 
studies will continue to shed light on the complex behaviour 
of these fascinating systems.
\end{abstract}
\noindent{\it Keywords:} Magic Number, Single-particle energy, 
Quadrupole Deformation, Nuclear Symmetry Energy, Coherent 
Density Fluctuation Model, Relativistic Mean-Field
\section{Introduction}
The current area of active research in nuclear structure studies is focused on 
the drip-line region of the nuclear landscape, both theoretically and 
experimentally. This region represents the boundary beyond which nuclei can no 
gamb90longer hold onto their extra nucleons (neutrons or protons), resulting in an 
unstable configuration. There exists a considerable difference in nuclear 
structure properties as one moves from the line of $\beta-$stability toward the 
drip lines \cite{doba94,wern94,chou95,ren94,gupt97,patr97}. For example, the 
sequence of magic numbers in the light mass region changes significantly as one 
moves from the stability line to the drip-line regions \cite{meht03}. In 
particular, the behaviour of magic numbers, which typically signify closed 
shells of nucleons, is not straightforward when approaching the drip lines. 
Theoretical predictions and experimental observations often diverge, especially 
in the context of nuclei beyond the $\beta-$stability line. A notable example 
is the calcium isotopes between $^{40}$Ca and $^{48}$Ca, where long-known 
features like a parabola-like shape in binding energies and strong odd-even 
staggering (OES) effects are observed \cite{mill19}. These effects are 
especially prominent and persist even further into the neutron-deficient 
region, offering valuable insights into the evolution of nuclear structure 
under extreme conditions. One particularly striking feature is the behaviour of 
charge radii in calcium isotopes. Beyond neutron number N = 29, the charge 
radii of calcium isotopes increase rapidly, with the radius of $^{52}$Ca being 
significantly larger than that of $^{48}$Ca. This observation is unexpected, as 
N = 32 is believed to be a magic number for calcium isotopes 
\cite{wien13,huck85}. The larger-than-expected radius of $^{52}$Ca raises 
important questions about the underlying nuclear structure and the stability of 
nuclei in this region. 

Understanding shell structure helps in explaining the fundamental nuclear 
properties and processes, from the stability of atomic nuclei to the dynamics 
of nuclear reactions. The shell structure may depend on isospin asymmetry more 
strongly than would be predicted from most conventional theories \cite{sorl08}. 
The new magic numbers, N = 16 and 32, have also been revealed in neutron-rich 
nuclei along with the elimination of N = 8 and 20 \cite{ozaw20,kanu02}. 
Experimentally, the disappearance of the $N = 20$ magic shell due to 
significant deformation in $^{32}$Mg was demonstrated in early studies 
involving $\beta$-decay and Coulomb excitation \cite{guill84,moto95}. 
Similarly, the loss of $N = 28$ as a magic number was initially theorized for 
nuclei such as $^{40}$Mg, $^{42}$Si, and $^{44}$S \cite{wern94,ren94,gupt97}, 
and later confirmed experimentally through measurements of large quadrupole 
deformations \cite{glas97}. Additionally, the mixing of $1s_{1/2}$ and 
$0p_{1/2}$ orbitals in the ground-state wave functions of neutron-rich 
$^{11}$Li and $^{12}$Be nuclei negates the magic character of $N = 8$ 
\cite{simo99,navi00}, with a notable presence of the $0d_{5/2}$ component 
expected in $^{12}$Be \cite{navi00}. Recent experiments on $^{52}$Ca have 
identified $N = 32$ as a new magic number \cite{kanu02,pris01}. Furthermore, 
shell model calculations suggest that the magic nature is even stronger at 
$N = 34$ due to the tensor force \cite{honm05}, though data from $^{56}$Ti do 
not support the magicity at $N = 34$ \cite{forn04}. While $N = 40$ exhibits 
characteristics of a magic number in $^{68}$Ni \cite{brod95}, predictions 
for $^{60}$Ca remain contradictory \cite{naka08,naka10,tera06}. Following 
traditional calculations, the $N = 34$ shell closure was not confirmed by the 
two-neutron shell gaps in Ti and V isotopes \cite{iimu23}. Extensive studies 
have investigated whether the $N = 40$ subshell closure is a local phenomenon 
restricted to the magic nickel chain \cite{malb22,groo20,babc16}. More findings 
indicate that the charge radii of copper isotopes show only a weak $N = 40$ 
subshell closure effect \cite{biss16}, and chromium isotopes exhibit a new 
island of inversion at $N = 40$ \cite{moug18}. For the traditional magic 
numbers 28 and 50, the first direct experimental evidence for the doubly-magic 
nature of $^{78}$Ni has been found, which also confirms a deformed second 
low-energy $2^+$ state, supporting the prediction of shape coexistence in 
$^{78}$Ni \cite{nowa16,tani19}. Additionally, the measurement of the charge 
radius of $^{56}$Ni directly supports its doubly magic nature \cite{somm22}.

While bulk properties such as binding energy and two-neutron separation energy are fundamental probes of nuclear structure, their interpretation can be challenging in nuclei far from the valley of $\beta$-stability. In these neutron-rich regions, the large neutron-proton asymmetry enhances the prominence of the isospin-dependent part of the nuclear force \cite{satp04}. This necessitates the use of observables that are explicitly sensitive to isospin asymmetry. Nuclear symmetry energy is a key quantity in this regard, as it quantifies the energy cost of converting symmetric nuclear matter into asymmetric nuclear matter. However, a conceptual challenge exists: symmetry energy is formally defined for infinite nuclear matter in momentum space, whereas we study finite nuclei in coordinate space. To bridge this gap, the coherent density fluctuation model (CDFM) provides a robust formalism \cite{gaid11,gaid12,bhuy18,prav22}. The CDFM relates the properties of a finite nucleus to those of infinite nuclear matter by considering the nucleus as a superposition of spherical droplets of varying density. This approach is powerful because the density dependence of the symmetry energy is directly linked to the isovector component of the nuclear interaction and physical observables like the neutron skin thickness \cite{myer80,dani06,gaid12,bhuy18}. Crucially, recent studies have shown that the symmetry energy and its components can serve as sensitive indicators of shell and sub-shell closures, offering a complementary perspective to traditional bulk property analysis, especially for nuclei near and beyond the drip-line \cite{gaid12,anto16,bhuy18,prav22,prav_25npa}.

The evolving landscape of nuclear shell structure, particularly the controversies surrounding N = 34 and 40, motivates a re-examination using complementary theoretical approaches. In this study, we investigate the isotopes of Cl, Ar, K, Ca, Sc, Ti, V, and Cr, to assess the persistence of sub-shell/shell closure. We first employ the axially deformed relativistic mean-field (RMF) model with the NL3 parameterisation to calculate bulk nuclear properties, and to complement this analysis, we then utilise the CDFM procedure to investigate surface properties, namely the symmetry energy and its components. This provides an independent probe of the shell structure through isospin-dependent surface properties, allowing for a more robust validation of the findings. The paper is structured as follows: Section \ref{ssec:theory_rmf} provides a brief overview of the RMF framework. Section \ref{ssec:method_cdfm} details the CDFM methodology used to study the isospin-dependent properties. The results of our calculations are presented and discussed in Section \ref{result}, and our conclusions are summarized in Section \ref{summary}.

\section{Theoretical Formalism}  
\label{sec:theory}
This section presents the theoretical framework used to explore the structure of finite nuclei, focusing on the connection between bulk nuclear matter properties and surface phenomena driven by isospin asymmetry. The core of our method is the Relativistic Mean-Field (RMF) theory, which offers a self-consistent and relativistic treatment of the nuclear many-body problem. Paired with the Bardeen-Cooper-Schrieffer (BCS) approach, RMF-BCS effectively describes the interior (bulk) properties of nuclei, including binding energies, density distributions, and single-particle levels, across various isotopic and isotonic chains. However, understanding surface effects, especially those influenced by siospin asymmetry, requires a complementary perspective. To this end, we incorporate the Coherent Density Fluctuation Model (CDFM), which bridges infinite nuclear matter calculations in momentum space with the spatially dependent properties of finite nuclei. CDFM treats the nucleus as a superposition of density profiles, enabling the study of observables such as the symmetry energy and neutron skin thickness. Together, the RMF-BCS and CDFM frameworks provide a comprehensive toolset for examining both the core and surface behavior of nuclei, with particular attention to shell structure and isospin-sensitive effects. Further technical details are discussed in the following subsections.

\subsection{Relativistic mean-field model} \label{ssec:theory_rmf}
In this calculation, we use the axially deformed relativistic mean-field (RMF) model, a powerful framework for accurately describing the ground-state properties of nuclei \cite{wale74,sero86,rein86,gamb90,rufa88,hard88,patr97,gupt97,patr09,bhuy12,bhuy15,bhuy18}. One of the key strengths of the RMF model is its natural inclusion of the spin-orbit interaction, which emerges from the meson-nucleon exchanges \cite{horo81}. This spin-orbit coupling is essential for correctly reproducing nuclear energy levels, particularly in deformed nuclei, and reflects the underlying relativistic dynamics of nucleon interactions within the nuclear medium. This model initiates with the Lagrangian density, represented as follows:
\begin{eqnarray}
\label{eqn1}
{\cal L} &=& \bar \psi_i(i\gamma^\mu\partial_\mu - M)\psi_i
+\frac{1}{2}\partial^\mu\sigma\partial_\mu\sigma - \frac{1}{2}m^2_\sigma\sigma^2 - \frac{1}{3}g_2\sigma^3- \frac{1}{4}g_3\sigma^4\nonumber\\ 
&-& g_s\bar\psi_i\psi_i\sigma - \frac{1}{4}\Omega^{\mu\nu}\Omega_{\mu\nu} 
+ \frac{1}{2}m^2_\omega V^\mu V_\mu
- g_\omega\bar\psi_i\gamma^\mu\psi_iV_\mu
- \frac{1}{4}\vec B^{\mu\nu}.\vec B_{\mu\nu}  \nonumber\\
&+& \frac{1}{2}m^2_\rho\vec R^\mu.\vec R_\mu
- g_\rho\bar\psi_i\gamma^\mu\vec\tau\psi_i.\vec R^\mu -\frac{1}{4}F^{\mu\nu}F_{\mu\nu} 
- e\bar\psi_i\gamma^\mu \frac{(1-\tau_{3})}{2}\psi_i A_\mu. 
\end{eqnarray}
Here, $\sigma$, $V^\mu$, $R^\mu$, and $A^\mu$ represent the fields for $\sigma$, $\omega$, $\rho$, and photons (electromagnetic field), respectively. The $\psi_i$ denote the Dirac spinor of the nucleons, with coupling constants for linear terms, namely $g_\sigma$, $g_\omega$, $g_\rho$, and $\frac{e}{4\pi}=\frac{1}{137}$ for $\sigma$-, $\omega$-, $\rho$-mesons, and photons, respectively. The Greek letter $\vec\tau$ ($\vec\tau_{3}$) signifies the Pauli isospin matrix (the third component of $\tau$) for the nucleon spinor ($\tau_3=-1$ for neutron and +1 for proton). $g_2$ and $g_3$ denote the coupling constants for the non-linear terms of the $\sigma$ meson. The quantities, $M$, $m_\sigma$, $m_\omega$, and $m_\rho$ represent the masses of the nucleons, $\sigma$, $\omega$, and $\rho$ mesons, respectively. The field tensors $\Omega^{\mu\nu}$, $R^{\mu\nu}$, and $F^{\mu\nu}$ corresponding to $\omega$-, $\rho$-mesons, and the electromagnetic field, respectively, as appearing in the Lagrangian, are defined as follows:
\begin{eqnarray}
\Omega^{\mu\nu}&=&\partial^\mu\omega^\nu-\partial^\nu\omega^\mu\nonumber\\
R^{\mu\nu}&=&\partial^\mu\vec\rho^\nu-\partial^\nu\vec\rho^\mu
-g_\rho (\vec R^\mu \times \vec R^\nu)\nonumber\\
F^{\mu\nu}&=&\partial^\mu A^\nu-\partial^\nu A^\mu.
\end{eqnarray}
The quantities marked with overhead arrows denote iso-vector properties. The 
tensor $R^{\mu\nu}$ involves a non-Abelian vector field. However, for 
simplicity, we approximate $R^{\mu\nu}$ 
as $\approx \partial^\mu\vec\rho^\nu-\partial^\nu\vec\rho^\mu$. The standard 
harmonic oscillator formula is used to estimate the centre of mass motion, 
$E_{c.m.}=\frac{3}{4}\left(41A^{-1/3}\right){MeV}$. The parameter $\beta_2$ is 
determined through the proton and neutron quadrupole moment as, $Q = Q_n + Q_p$.
The quadrupole moment ($Q_p$) for proton is:
\begin{equation}
Q_p = \sqrt{\frac{16\pi}{5}} \int \rho_p(\mathbf{r}) r^2 Y_{20}(\theta) \, d^3r,
\end{equation}
 and similarly the quadrupole moment ($Q_n$) for neutron is:
\begin{equation}
Q_n = \sqrt{\frac{16\pi}{5}} \int \rho_n(\mathbf{r}) r^2 Y_{20}(\theta) \, d^3r
\end{equation}
Here, $\rho_p$(r) and $\rho_n$(r) are proton and neutron
density distributions, respectively and $Y_{20}$ is the
spherical harmonic function with $l = 2$ and $m = 0$.
Therefore, the total quadrupole moment is :
\begin{equation}
Q = Q_n + Q_p = \sqrt\frac{16\pi}{5}\frac{3}{4\pi}AR_0^2\beta_2
\label{quad}
\end{equation}
In Eq. (\ref{quad}), if we replace $A$ by $Z$ we get the 
quadrupole moment for proton ($Q_p$), and if we replace $A$ 
by $N$ then we get the quadrupole moment for neutron ($Q_n$) with corresponding 
$\beta_2$ values for proton and neutron. And, the root mean square (rms) radius 
is, $\left\langle r_m^2\right\rangle = 
\frac{1}{A}\int\rho\left(r_{\perp},z\right)r^2d\tau$, 
where $\rho\left(r_{\perp},z\right)$ is the axially deformed density and $A$ 
is the mass number. Using the conventional relations provided in Ref. 
\cite{pann87}, it is also possible to derive the total binding energy and 
other observables.
For nuclei situated relatively close to the $\beta$-stability line, employing 
the constant gap BCS pairing approach can yield a reasonably accurate 
approximation of pairing \cite{doba84}. We are using the constant gap 
Bardeen-Cooper-Schrieffer (BCS) technique with the NL3 parameter set in the 
current analysis  \cite{madl88,moll88,bhuy15,type01,patr09}. In other words, 
the constant gap BCS technique can be used to handle the current mass region 
without causing too many difficulties in calculations 
\cite{patr09, madl88, moll88}. The pairing energy is expressed  in terms of 
occupation probability $v_i^2+u_i^2=1$\cite{gamb90,patr93,pres82} as:
\begin{equation}
    E_{\text {pair}}=-G\left[\sum_{i>0} u_i v_i\right]^2
\end{equation}
where $G$ is the pairing force constant. The BCS equation 
$2 \epsilon_i u_i v_i-\Delta\left(u_i^2-v_i^2\right)=0$ is obtained via the 
variational approach of the occupation numbers $v_i^2$ \cite{pres82} and the 
pairing gap is defined by,
\begin{equation}
\Delta=G \sum_{i>0} u_i v_i
\end{equation}
For pairing energy, this is the well-known BCS equation with the occupation 
number, 
\begin{equation}  
n_i=v_i^2=\frac{1}{2}\left[1-\frac{\epsilon_i-\lambda}{\sqrt{\left(\epsilon_i-\lambda\right)^2+\Delta^2}}\right]. 
\end{equation}
The pairing correlations will be taken into account in the constant gap 
approximation taken from the prescription of Madland and Nix \cite{madl88}. Here, $\Delta_p=R B_s e^{s I-tI^2} / Z^{1/3}$ and  $\Delta_n=R B_s e^{-sI-tI^2} / A^{1/3}$ are the conventional expressions for the pairing gaps of the proton and neutron, respectively, where, $I=(N-Z)/(N+Z)$, $R=5.72, s=0.118, t=8.12, B_s=1$, these gaps are true for nuclei in this mass region that are on and off the stability line. The occupation probability and chemical potentials $\lambda_n$ and $\lambda_p$ from the particle numbers are computed using the preceding equations and the gap parameter. The energy of
 pairing by using  $\Delta_p$ and $\Delta_n$ as,
\begin{equation}
E_{pair} =-\frac{\Delta^2}{G}= -\Delta\sum_{i>0}u_i v_i
\end{equation}
For a constant pairing gap, the pairing energy should vary with particle number since it is dependent upon the occupancy probability $v_i^2$ and $u_i^2$. It is commonly known that for a constant pairing gap $\Delta$ and force constant $G$, the pairing energy $E_{pair}$ diverges if it is extended to an infinite configuration space. For an odd nucleon system, the blocking approximations are taken into account, and details can be found in Refs. \cite{patr01,meht02,saho20}.
\subsection{Coherent density fluctuation model}
\label{ssec:method_cdfm}
The theoretical framework for this study is the coherent density fluctuation model (CDFM). The CDFM originates from the generator coordinate method (GCM) in its $\delta$-function limit \cite{anto80, anto79, anto82, gaid11, gaid12, grif57}. The central idea of the model is that the nucleus is not a static object but can be described as a coherent superposition of spherical, uniform-density nuclear matter configurations known as ``fluctons''. The model's cornerstone is the one-body density matrix (OBDM), $\rho(\textbf{r}, \textbf{r'})$, which is constructed by integrating over the OBDMs of all possible fluctons, $\rho_{x}(\textbf{r}, \textbf{r'})$:
\begin{equation}
\label{eqn:1}
\rho(\textbf{r},\textbf{r}^{\prime}) = \int_{0}^{\infty} |\mathcal{F}(x)|^{2} \rho_{x}(\textbf{r},\textbf{r}^{\prime}) dx.
\end{equation}
In this expression, $\rho_{x}$ represents the OBDM for a system of $A$ nucleons confined within a spherical flucton of radius $x$. The function $|\mathcal{F}(x)|^{2}$ is the weight function, which dictates the probability distribution of finding a flucton of a given size $x$. This weight function is normalized such that $\int_{0}^{\infty} |\mathcal{F}(x)|^{2} dx = 1$. For a flucton of radius $x$, the OBDM is given by the standard expression for uniform nuclear matter:
\begin{equation}
\label{eqn:2}
\rho_x(\mathbf{r},\mathbf{r}^\prime) = 3\rho_0(x)\frac{J_1(k_F(x)|\mathbf{r}-\mathbf{r}^\prime|)}{k_F(x)|\mathbf{r}-\mathbf{r}^\prime|},
\end{equation}
where $J_1$ is the first-order spherical Bessel function. The density of the flucton is $\rho_{o}(x) = 3A/(4\pi x^{3})$, and $k_F(x)$ is the corresponding Fermi momentum for the nucleons, defined as:
\begin{equation}
\label{eqn:3}
k_{F}(x) = \left(\frac{3\pi^{2}}{2}\rho_{o}(x)\right)^{1/3} = \left(\frac{9\pi A}{8x^3}\right)^{1/3}.
\end{equation} 
To determine the weight function $|\mathcal{F}(x)|^{2}$, one could solve a complex differential equation derived from the GCM \cite{anto79,anto82, anto94}. However, a more direct and widely used method relates $|\mathcal{F}(x)|^{2}$ to the slope of the nucleus's ground-state density profile $\rho(r)$ \cite{anto82,anto94}. This is the approach we adopt:
\begin{equation}
\label{eqn:21}
|\mathcal{F}(x)|^{2} = -\frac{1}{\rho_{o}(x)} \frac{d\rho(r)}{dr} \bigg|_{r=x}.
\end{equation}
This formulation ensures that the total mass number is correctly normalized, 
i.e., $\int \rho(\mathbf{r}) d\mathbf{r} = A$.
\subsubsection{Calculation of nuclear symmetry energy}
\label{sec:theory_Sv_Ss}
Having established the CDFM framework, we apply it to calculate the nuclear symmetry energy, $S$. Within the model, $S$ for a finite nucleus is an average of the infinite nuclear matter symmetry energy, $S^{\mathrm{NM}}(x)$, weighted by the flucton distribution \cite{anto94, gaid11, gaid12, sarr07, bhuy18}:
\begin{equation}
\label{eqn:Sf}
S = \int_{0}^{\infty} dx |\mathcal{F}(x)|^{2} S^{\mathrm{NM}}(x).
\end{equation}
The term $S^{\mathrm{NM}}(x)$ is the symmetry energy at the local density $\rho_o(x)$ of a flucton with radius $x$. For this quantity, we employ the well-established Brueckner energy density functional (B-EDF) \cite{brue68,brue69}. According to the B-EDF, the energy per nucleon in asymmetric nuclear matter (with asymmetry $\delta = (N-Z)/A$) is:
\begin{align}
	\frac{E}{A}(\rho_0, \delta) &= 37.53\left[(1+\delta)^{5/3}+(1-\delta)^{5/3}\right]\rho_0^{2/3} + b_1 \rho_0 + b_2 \rho_0^{4/3} + b_3 \rho_0^{5/3} \nonumber \\
	&+ \delta^2\left[b_4 \rho_0 + b_5 \rho_0^{4 / 3} + b_6 \rho_0^{5/3} \right].
\end{align}
The parameters used are $b_1=-741.28$, $b_2=1179.89$, $b_3=-467.54$, $b_4=148.26$, $b_5=372.84$, and $b_6=-769.57$. The nuclear symmetry energy is defined as the coefficient of the $\delta^2$ term in the expansion of $E/A$. By expanding the kinetic term and collecting all $\delta^2$ coefficients, one obtains the expression for $S^{\mathrm{NM}}$ at density $\rho_0(x)$:
\begin{equation}
\label{symfit}
S^{\mathrm{NM}}(x) = 41.7\rho_0^{2/3}(x)+b_4\rho_0(x) + b_5\rho_0^{4/3}(x)+b_6\rho_0^{5/3}(x).
\end{equation} 
To further analyze the calculated symmetry energy, we use the Danielewicz prescription \cite{dani09}, which decomposes $S$ into volume ($S_V$) and surface ($S_S$) contributions:
\begin{equation}
\label{eqn:23}
S = \frac{S_{V}}{1+\frac{S_{S}}{S_{V}}A^{-1/3}} = \frac{S_{V}}{1+\frac{1}{\kappa A^{1/3}}},
\end{equation}
where $\kappa = S_V / S_S$ is the ratio of the volume to surface symmetry energies. This ratio can be calculated directly within the CDFM formalism as follows \cite{anto16,danc20}:
\begin{equation}
\label{eqn:19_cdfm}
\kappa = \frac{3}{r_{0} \rho_{0}} \int_{0}^{\infty} dx |\mathcal{F}(x)|^{2} x \rho_{o}(x) \left[\frac{S^{\mathrm{NM}}(\rho_{o})}{S^{\mathrm{NM}}(x)}-1\right].
\end{equation}
Here, $\rho_{0}$ is the equilibrium nuclear matter density, $S^{\mathrm{NM}}(\rho_{o})$ is the symmetry energy at that density, and $r_{0}$ is the corresponding nuclear radius constant \cite{diep07,anto16}. Once the nuclear symmetry energy $S$ (Eq.~\ref{eqn:Sf}) and the ratio $\kappa$ (Eq.~\ref{eqn:19_cdfm}) are computed, the volume and surface components are readily extracted by rearranging Eq.~\eqref{eqn:23} \cite{anto16,prav_25npa}:
\begin{equation}
\label{eqn:Sv}
S_{V} = S \left(1+\frac{1}{\kappa A^{1/3}}\right),
\end{equation}
and    
\begin{equation}
\label{eqn:Ss}
S_{S} = \frac{S_V}{\kappa} = \frac{S}{\kappa} \left(1+\frac{1}{\kappa A^{1/3}}\right).
\end{equation}
\begin{figure}[b]
\begin{center}
\includegraphics[width=12.0 cm, height=8.0cm, clip=true]{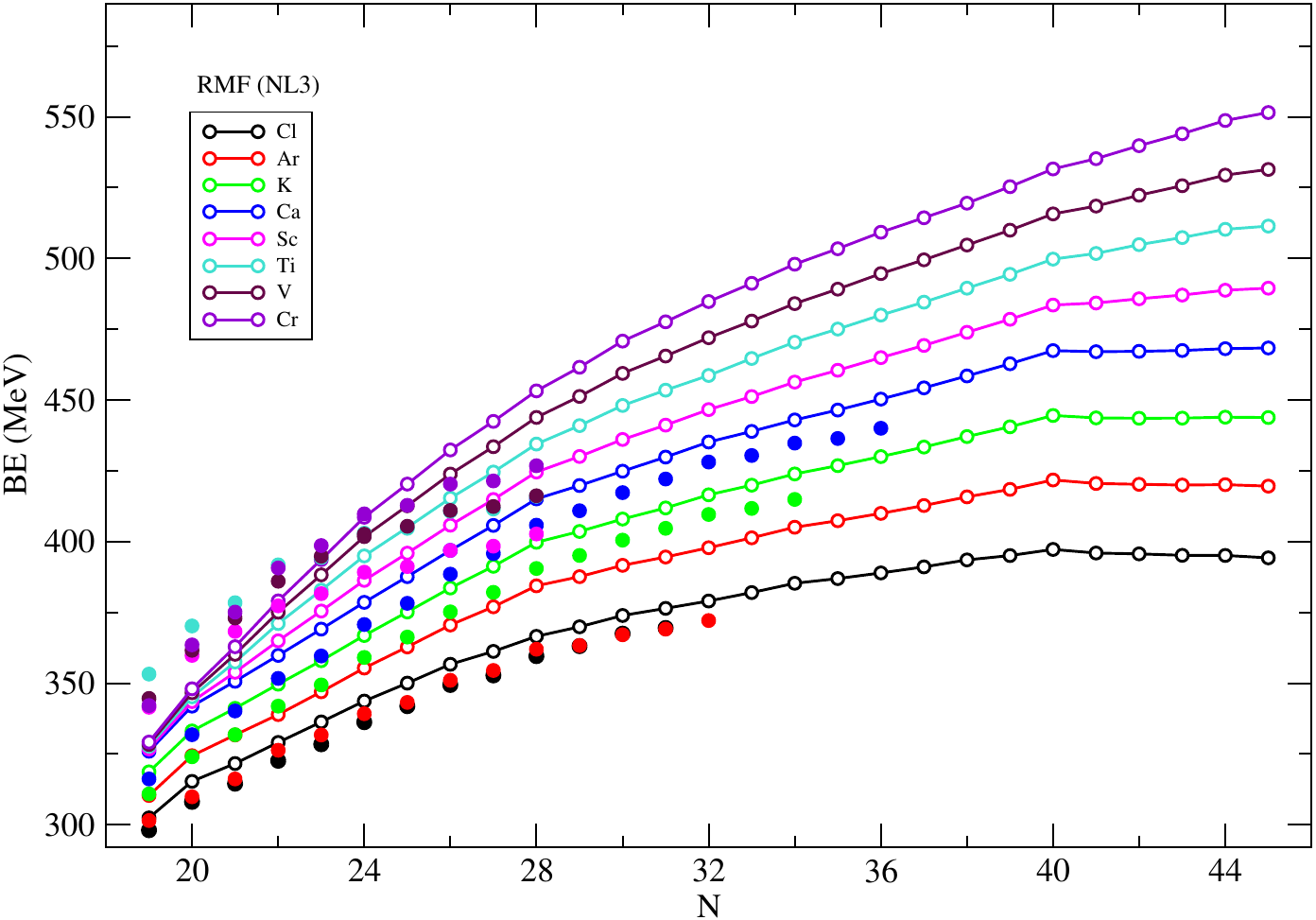}
\end{center}
\caption{The binding energy as a function of neutron number using RMF with NL3 parameter set for the isotopic chain of Cl-Cr nuclei compared with the available experimental data \cite{wang21,prit16}. The solid circle represent the corresponding experimental data.}
\label{fig1}
\end{figure}
\begin{figure}[t]
\begin{center}
\includegraphics[width=12cm,height=8cm]{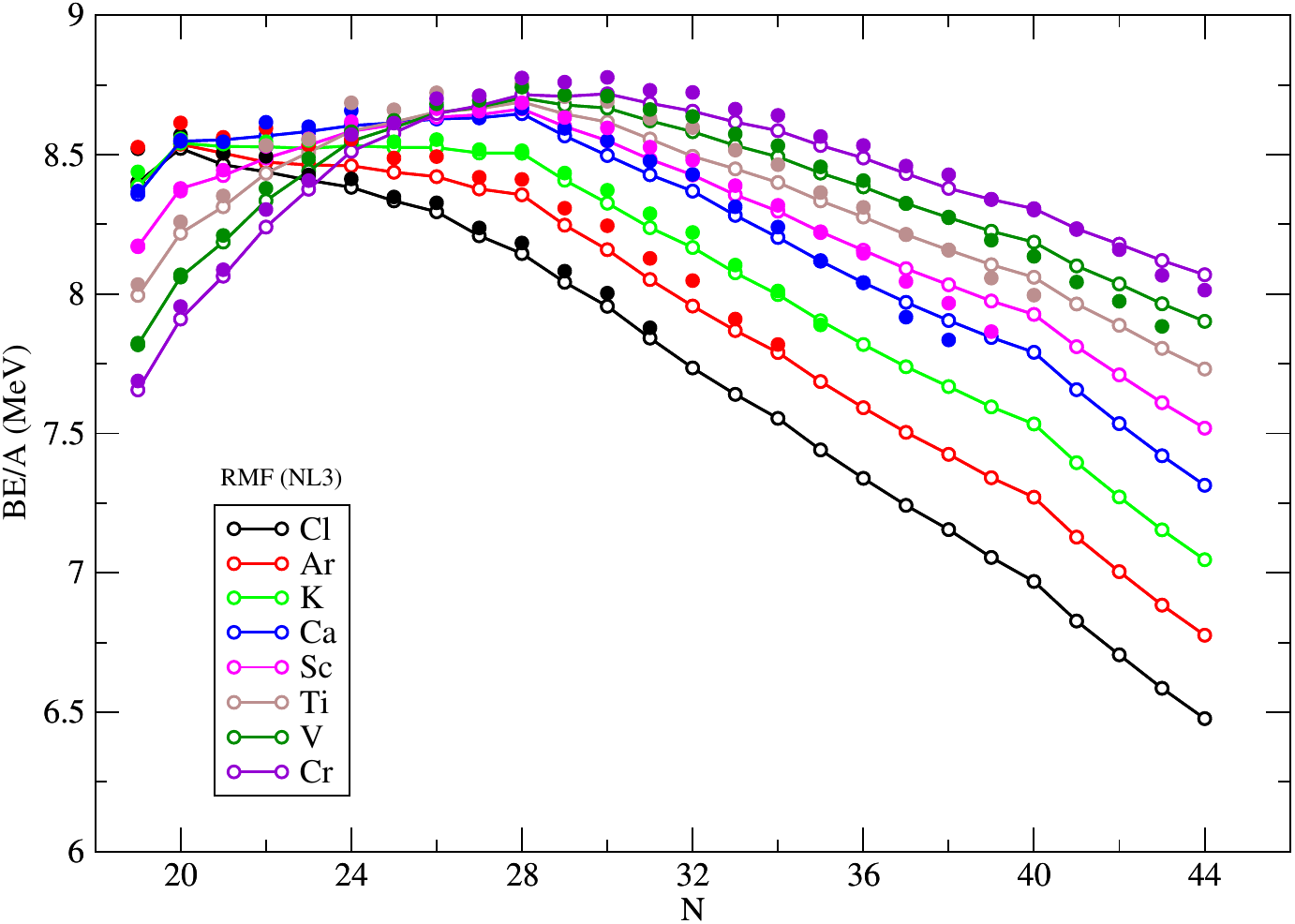}
\end{center}
\caption{The binding energy as a function of neutron number from RMF with NL3 parameter set for the isotopic chain of Cl - Cr nuclei compared with the available experimental data \cite{wang21,prit16}. The solid circle represent the corresponding experimental data.}
\label{fig2}
\end{figure}
\section{Results and Discussion} \label{result}
To investigate the ground-state properties of neutron-rich isotopes, we analyse binding energy, binding energy per nucleon, charge radii, two-neutron separation energies, quadrupole deformation parameters, and single-particle energy levels for isotopic chains with Z = 17 to Z = 24. These calculations are performed using the axially deformed Relativistic Mean-Field (RMF) model with the NL3 parameter set. Specifically, we examine the isotopes of Chlorine (Cl), Argon (Ar), Potassium (K), Calcium (Ca), Scandium (Sc), Titanium (Ti), Vanadium (V), and Chromium (Cr). The shell numbers are set as  $N_F$ = 12 for fermions and $ N_B$ = 12 for bosons, by testing the convergence solution for this mass region. As the goal of this study is to identify potential sub-shell or shell closures in neutron-rich nuclei, we extend our analysis to include isospin-dependent observables—specifically the symmetry energy and its components—using the coherent density fluctuation model (CDFM), with input derived from the RMF-based ground-state properties.
\subsection{Nuclear Binding Energy}
In the present study, we calculate the total binding energy and binding energy per nucleon for the isotopic chains of Cl, Ar, K, Ca, Sc, Ti, V, and Cr, as depicted in Figs. \ref{fig1}–\ref{fig2}. These results are systematically compared with experimental data from Refs. \cite{wang21,prit16}. Overall, the RMF (NL3) model reproduces the experimental trends well, with particularly good agreement observed for the Calcium (Ca) and Titanium (Ti) isotopes. However, notable deviations appear in lighter nuclei such as Chlorine (Cl) and Argon (Ar), where the RMF predictions tend to slightly overestimate the binding energies, especially at higher neutron numbers. This overestimation may be attributed to the rapid shape transitions that occur in light nuclei due to small changes in neutron number, causing fluctuations between spherical and deformed configurations. Additionally, lighter nuclei are more susceptible to shell and sub-shell effects, which are not always fully captured in mean-field approximations. Despite these discrepancies, the binding energy curves for all isotopic chains are nearly parallel, indicating a consistent and reliable performance of the RMF model across the nuclear landscape considered in this work.

Fig. \ref{fig2} illustrates the binding energy per nucleon (B.E./A), providing insight into how tightly bound each nucleon is within the nucleus across the isotopic chains. As seen in the figure, B.E./A generally increases with neutron number, peaks around a specific neutron configuration, and then begins to decline for heavier isotopes. This trend reflects the well-known pattern of nuclear stability, where the most stable configurations—marked by maximal binding per nucleon—typically occur near closed-shell or magic numbers. In the current study, the Calcium (Ca) and Potassium (K) isotopes exhibit particularly high B.E./A values near mass numbers A = 40 and 48, corresponding to the magic numbers Z = 20 (N = 20), and Z = 20 (N = 28). These peaks in binding energy per nucleon indicate enhanced nuclear stability due to shell closures. In contrast, the Titanium (Ti), Vanadium (V), and Chromium (Cr) isotopes show a peak in stability around A = 48, consistent with the shell closure at N = 28. Interestingly, in isotopes like Ca, Sc, and Ti, the B.E./A values do not decline significantly beyond N = 28; instead, they show a plateau or a slight rise around N = 40. This deviation from the typical downward trend seen beyond the mid-shell region suggests additional structural stability in this region. Such behaviour aligns with the predictions of a sub-shell closure at N = 40, often considered a semi-magic number due to emerging evidence from both theoretical and experimental studies \cite{adri20}. The enhanced stability near N = 40 is further supported by the two-neutron separation energy, discussed in the subsequent subsection, which also reflects underlying shell effects in these neutron-rich isotopes.
\begin{figure}[t]
\begin{center}
\includegraphics[width=12cm,height=8cm]{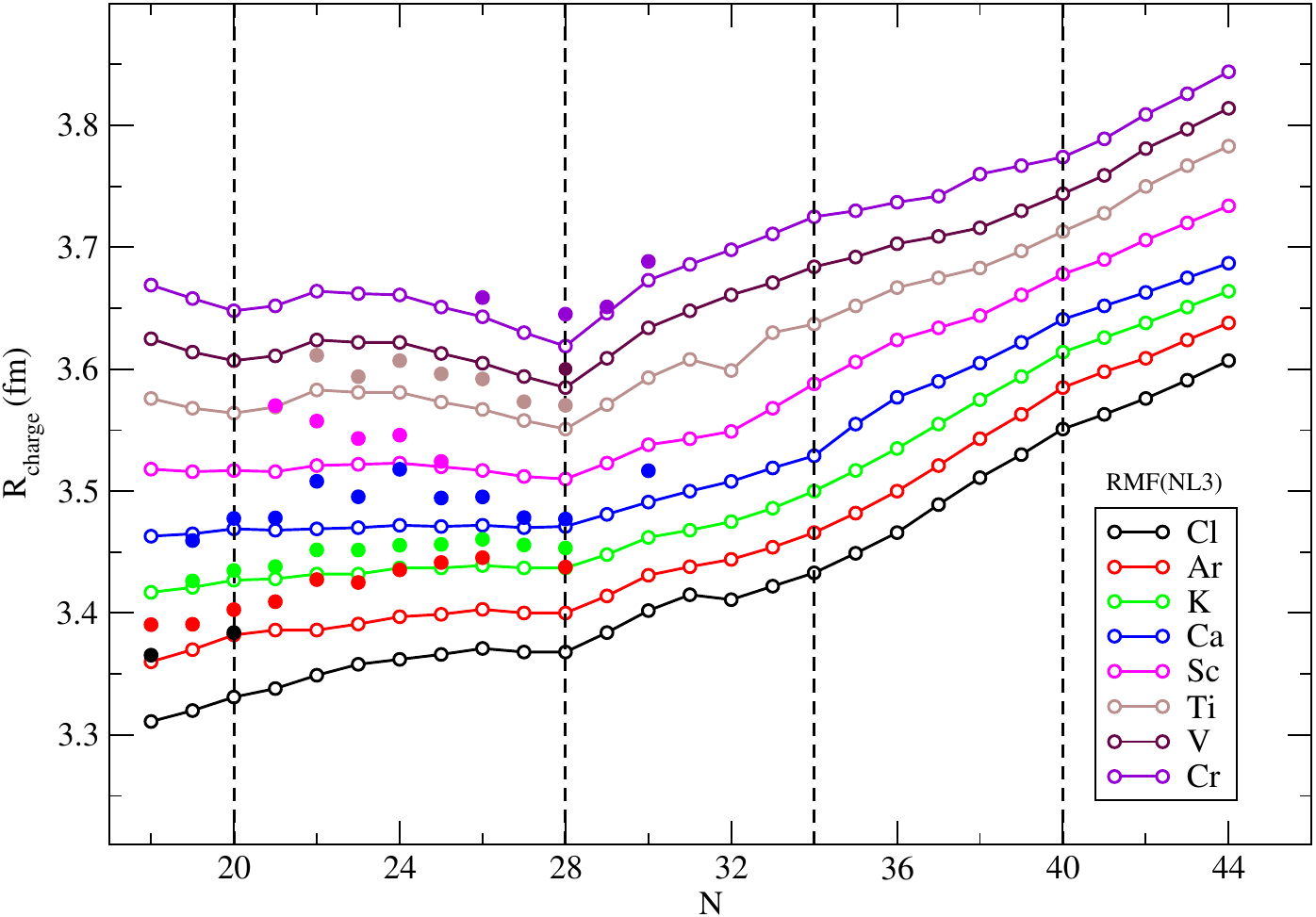}
\end{center}
\caption{The Charge radius as a function of Neutron number from RMF with NL3 parameter for the isotopic series of Cl, Ar, K, Ca, Sc, Ti, V, and Cr nuclei. }
\label{fig3}
\end{figure}
\subsection{Nuclear Charge Radius}
The charge radius of atomic nuclei is a key observable for probing nuclear structure, especially regarding how nuclei evolve with the addition of neutrons. It reflects the spatial distribution of protons and provides crucial insights into shell effects, deformation, and changes in nuclear density. As shown in Fig. \ref{fig3}, the charge radii of isotopes in the Cl, Ar, K, Ca, Sc, Ti, V, and Cr chains reveal distinct patterns that align with known and emerging magic numbers. A marked stabilisation in the growth of the charge radius is observed at neutron numbers N = 20 and N = 28, particularly in Ar, K, Ca, and Sc around N = 20, and in Ti, V, and Cr near N = 28. These plateaus correspond to closed-shell configurations, where the nuclear structure becomes more compact and resistant to deformation, signalling enhanced stability. Beyond these traditional magic numbers, subtle but significant variations occur at N = 34 and N = 40, suggesting possible sub-shell closures. For example, Cl, Ti, V, and Cr isotopes display a reduced rate of increase in charge radius around N = 34, indicating a temporary stabilization and hinting at sub-shell effects. A similar trend is seen around N = 40, where several nuclei show a flattening in the charge radius curve, pointing to increased resistance to spatial expansion and nuclear softening. These behaviours reinforce the idea that N = 34 and N = 40 may represent regions of enhanced stability due to evolving shell structures, particularly in neutron-rich systems. Thus, the analysis of charge radii not only supports established shell closures but also highlights regions of emerging magicity, contributing to a more nuanced understanding of nuclear compactness and the role of shell evolution in determining nuclear size and shape. The available experimental charge radii of 
the nuclei is shown in solid spheres \cite{ange13}.
\begin{figure}[t]
\begin{center}
\includegraphics[width=12cm,height=8cm]{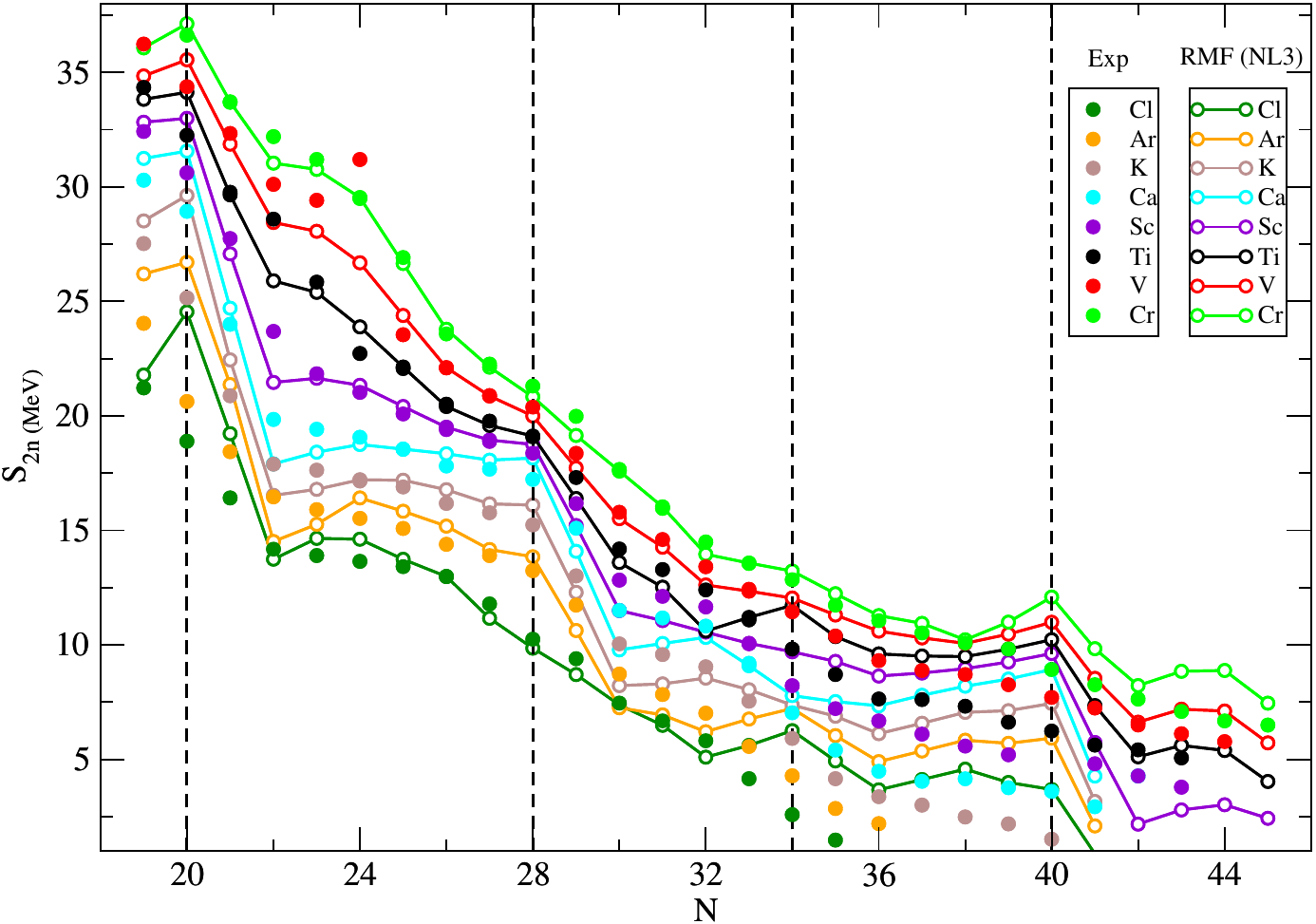}
\end{center}
\caption{The two-neutron separation energy for Cl, Ar, K, Ca, Sc, Ti, V and Cr nuclei from RMF (NL3) are compared with the experimental data\cite{wang21} }
\label{fig4}
\end{figure}
\subsection{Separation Energy}
The separation energies for two neutrons are calculated by the difference in binding energy of two isotopes using relations:
\begin{eqnarray}
S_{2n}(N,Z) = BE(N,Z) - BE(N-2,Z)\nonumber\\  
\end{eqnarray}
The two-neutron separation energy ($S_{2n}$) is a critical indicator of nuclear stability, as it reflects the energy required to remove two neutrons from a nucleus. A sudden drop or noticeable change in the slope of $S_{2n}$ across an isotopic chain often signals shell or sub-shell closures. Figure \ref{fig4} presents the calculated $S_{2n}$ values for the isotopes of Cl, Ar, K, Ca, Sc, Ti, V, and Cr, along with available experimental data. The strong agreement between theory and experiment confirms the reliability of the RMF model in capturing key structural features of these nuclei. Clear discontinuities or kinks are observed in the separation energy curves at neutron numbers, N=20, N=28, N=34, and N=40, which correspond to enhanced nuclear stability—hallmarks of magic numbers. Peaks in the $S_{2n}$ values at N=20 and N=28 validate these well-established shell closures, particularly in nuclei such as Ca and Ti. Interestingly, while N=34 is not traditionally classified as a magic number, the calculated $S_{2n}$ values for Cl, Ar, and Ti exhibit a local stabilisation or slower decrease in this region, suggesting the presence of a sub-shell closure. Similarly, the emergence of a peak in the separation energy at N=40 across multiple isotopic chains points to a possible shell closure, supporting the growing body of evidence for the semi-magic nature of N=40. The consistency of higher $S_{2n}$ values at these neutron numbers—both in theoretical predictions and experimental observations—reinforces the link between neutron shell closures and nuclear stability. These findings further support the existence of evolving shell structures in neutron-rich nuclei and highlight the utility of $S_{2n}$ as a sensitive probe for identifying magic and semi-magic numbers in the nuclear chart.
\begin{figure}[t]
\begin{center}
\includegraphics[width=13cm,height=9cm]{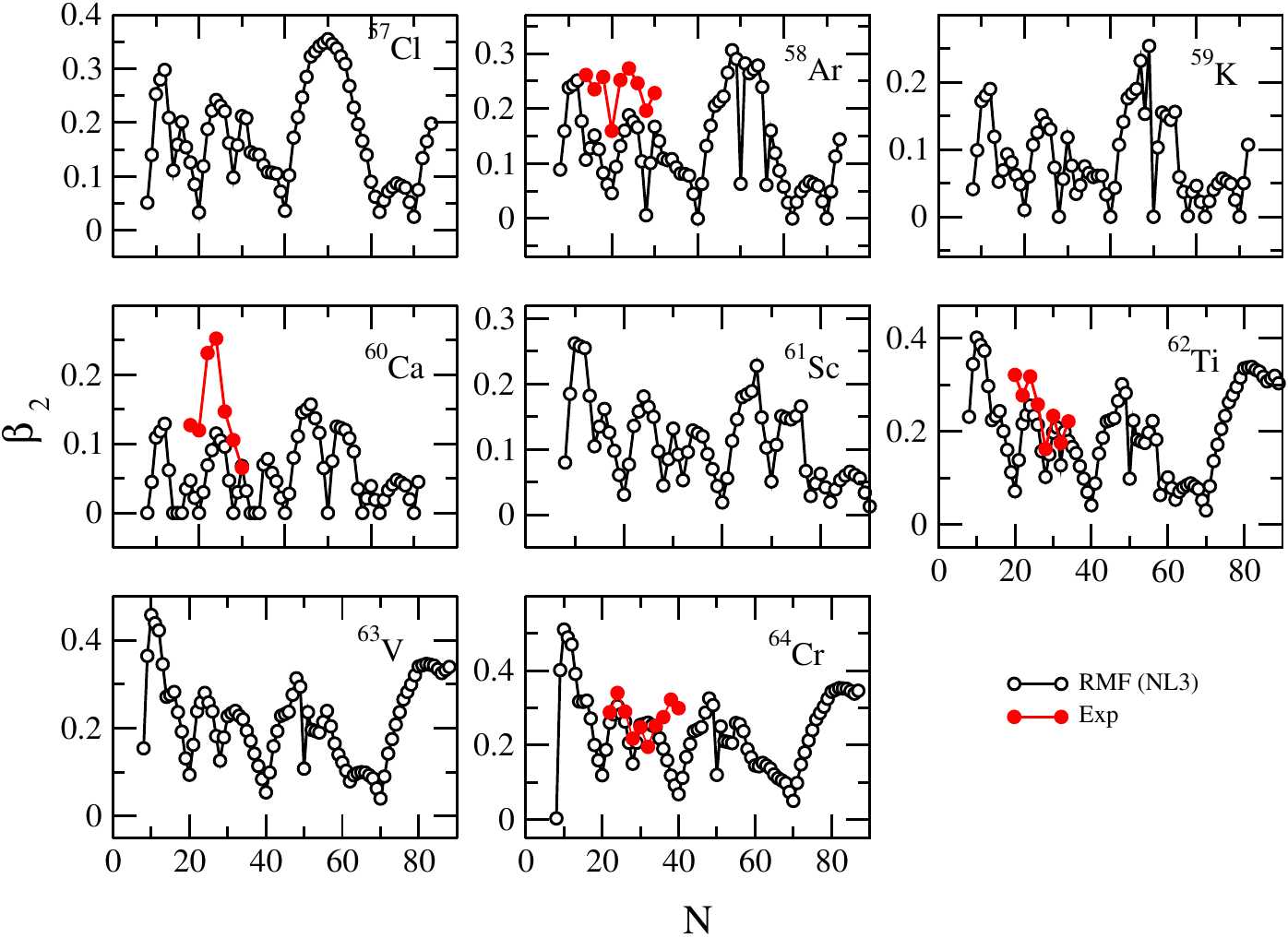}
\end{center}
\caption{The Quadrupole deformation of Cl, Ar, K, Ca, Sc, Ti ,V and Cr nuclei are compared with experimental data \cite{prit16}.}
\label{fig5}
\end{figure}
\subsection{Quadrupole Deformation Parameter}
The quadrupole deformation parameter ($\beta_2$) offers vital information about the shape evolution of nuclei as a function of neutron number, distinguishing between spherical, prolate (elongated), and oblate (flattened) configurations. In this work, we analyse the deformation behaviour of isotopes from Cl (Z = 17) to Cr (Z = 24) using the Relativistic Mean Field (RMF) model with the NL3 parameter set, comparing our results with experimental data where available. For Chlorine (Cl) isotopes, $\beta_2$ remains small at lower neutron numbers, indicating near-spherical shapes, but starts to increase around N=20, suggesting the onset of deformation—likely due to shell structure changes and weakening of the N = 20 magicity in light nuclei. The RMF predictions capture this trend well, with minor deviations attributed to shell and pairing effects. Argon (Ar) and Potassium (K) isotopes similarly exhibit low deformation near closed neutron shells, particularly at N = 20, supporting the presence of spherical or weakly deformed ground states. However, deformation increases beyond this point, especially in K isotopes, as the addition of neutrons leads to gradual shape changes. The RMF model reproduces these behaviours, showing a smooth increase in $\beta_2$ consistent with experimental observations. Calcium (Ca) isotopes display minimal deformation at N = 20, consistent with the double magic nature of $^{40}$Ca, but a noticeable increase in $\beta_2$ occurs around N = 28 and even more around N = 40, hinting at shape transitions influenced by sub-shell closure effects. This evolution is well reproduced by RMF calculations.

Scandium (Sc) and Titanium (Ti) isotopes show a more pronounced deformation development. For Sc, the increase in $\beta_2$ begins around N = 24, reflecting a departure from spherical symmetry, while Ti isotopes demonstrate a smoother and more consistent increase in deformation with neutron number, peaking near N=28 and continuing toward N=40, signalling the growing importance of deformation-driving orbitals. In Vanadium (V) and Chromium (Cr) isotopes, deformation becomes even more significant. RMF predictions show a steadily rising $\beta_2$ value beyond N=28, with Cr isotopes in particular displaying strong prolate deformation past this point. This trend reflects the development of a deformed region—often referred to as an “island of inversion”, centred around N=40, where traditional shell closures weaken, and collective effects dominate. Overall, the RMF (NL3) model effectively captures the general trends in nuclear deformation across these isotopic chains. Notably, near traditional and emerging magic numbers (e.g., N = 20, 28, 34, and 40), abrupt changes or plateaus in $\beta_2$ indicate increased stability and shell effects. While the theoretical framework shows good agreement with available experimental data, further refinements—such as including beyond-mean-field correlations and shape coexistence—may be required to fully capture subtle structural features in transitional nuclei.
\begin{figure}
\centering
\includegraphics[width=12cm,height=8cm]{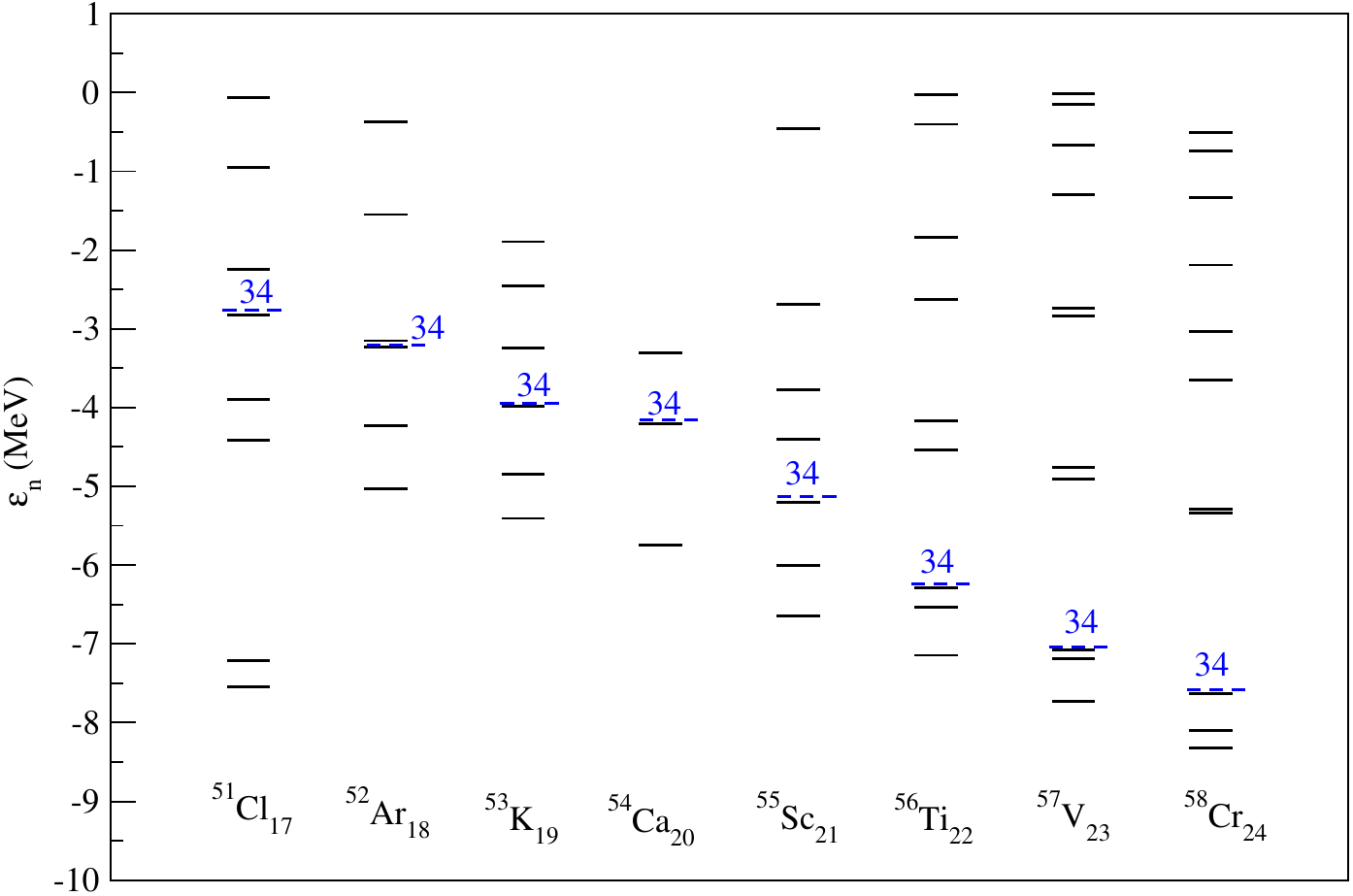}
\caption{ The neutron single-particle energy of Cl, Ar, K, Ca, Sc, Ti, V and Cr nuclei to examine the shell gap at N = 34 within the relativistic mean field approach.}
\label{fig6}
\end{figure}
\begin{figure}
\centering
\includegraphics[width=12.0cm,height=8.0 cm]{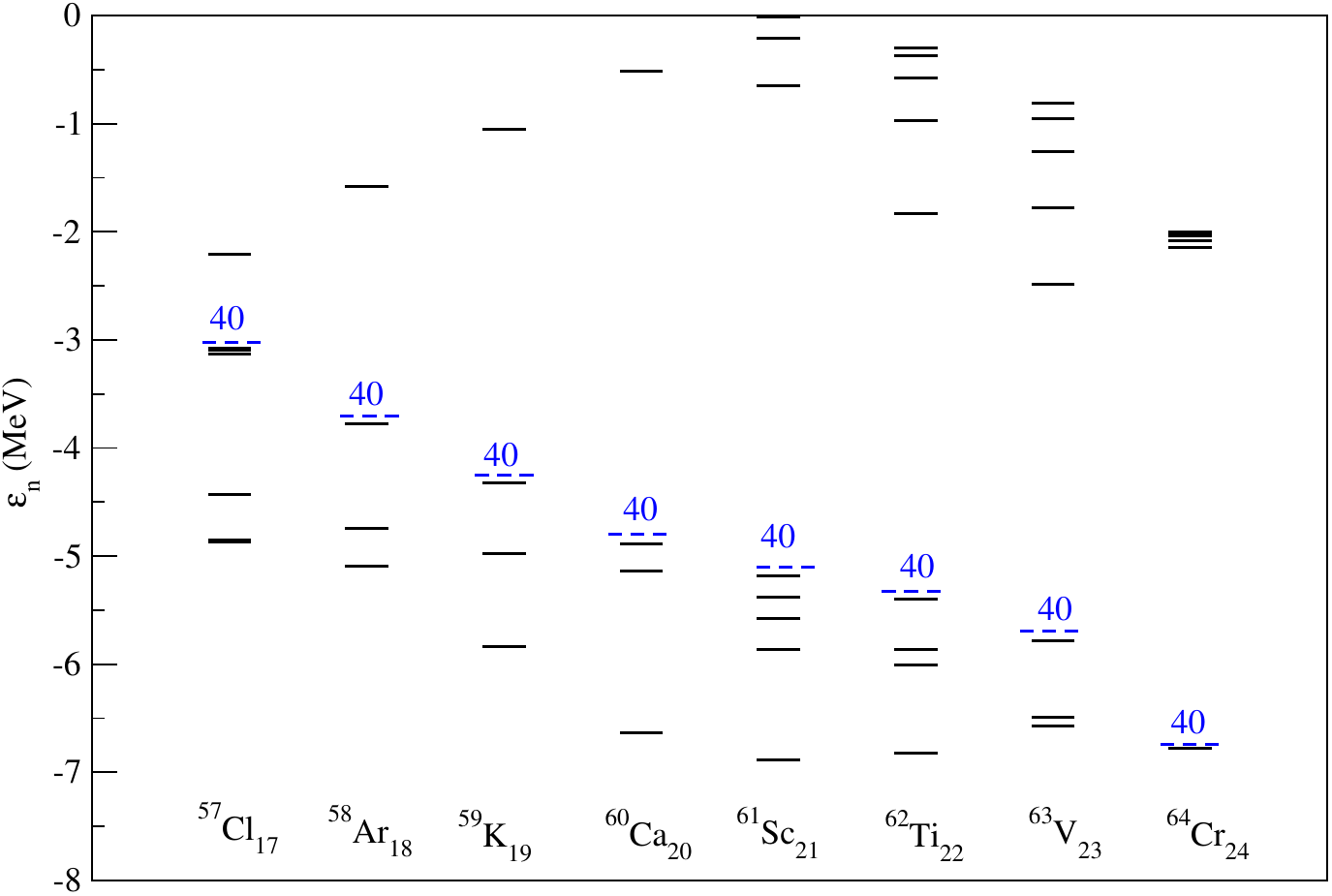}
\caption{The neutron single-particle energy of Cl, Ar, K, Ca, Sc, Ti, V and Cr nuclei to examine the shell gap at N = 40 within the relativistic mean field approach.}
\label{fig7}
\end{figure}
\subsection{Single Particle Energy}
The single-particle energy levels, as shown in Figs. 6 and 7, offer critical insight into the shell structure of neutron-rich nuclei ranging from Chlorine (Cl) to Chromium (Cr). These energy levels help identify shell or sub-shell closures by revealing gaps between occupied and unoccupied orbitals. In Fig. \ref{fig6}, a distinct shell gap is observed at the neutron number N = 34, though the magnitude varies across different isotopes. For lighter nuclei such as $^{51}$Cl, $^{52}$Ar, $^{53}$K, $^{54}$Ca, and $^{55}$Sc, the shell gap is present but relatively modest. However, as we move to heavier isotopes $^{56}$Ti, $^{57}$V, and $^{58}$Cr, the gap becomes more pronounced, indicating stronger shell effects. The presence of a consistent energy gap at N=34 across these isotopic chains suggests the emergence of a sub-shell closure, especially in the mid-mass region. While N = 34 is not traditionally classified as a magic number, the observed energy separation supports the idea that it behaves as a magic number in neutron-rich systems. The systematic increase in the gap's magnitude with proton number indicates an evolving shell structure, likely influenced by changes in the spin-orbit interaction and tensor forces, as also suggested by modern shell model calculations.

Fig. \ref{fig7} further highlights a substantial energy gap at N = 40, particularly in nuclei such as $^{60}$Ca and $^{61}$Sc. Here, the single-particle energy diagram reveals a sharp drop between the highest occupied and the lowest unoccupied neutron orbitals, characteristic of a shell closure. This large energy gap at N = 40 mirrors the classical magic numbers in terms of its stabilising effect and reflects an emerging closed-shell behaviour, especially in neutron-rich environments. Together, these findings validate theoretical predictions from the RMF model and reinforce the importance of single-particle level studies in identifying magic and sub-magic numbers. The consistent appearance of energy gaps at N = 34 and N = 40 across several nuclei strengthens the case for their magicity and highlights their relevance in refining nuclear shell models, particularly for isotopes far from stability.
\begin{figure}
\centering
\includegraphics[width=0.9\columnwidth]{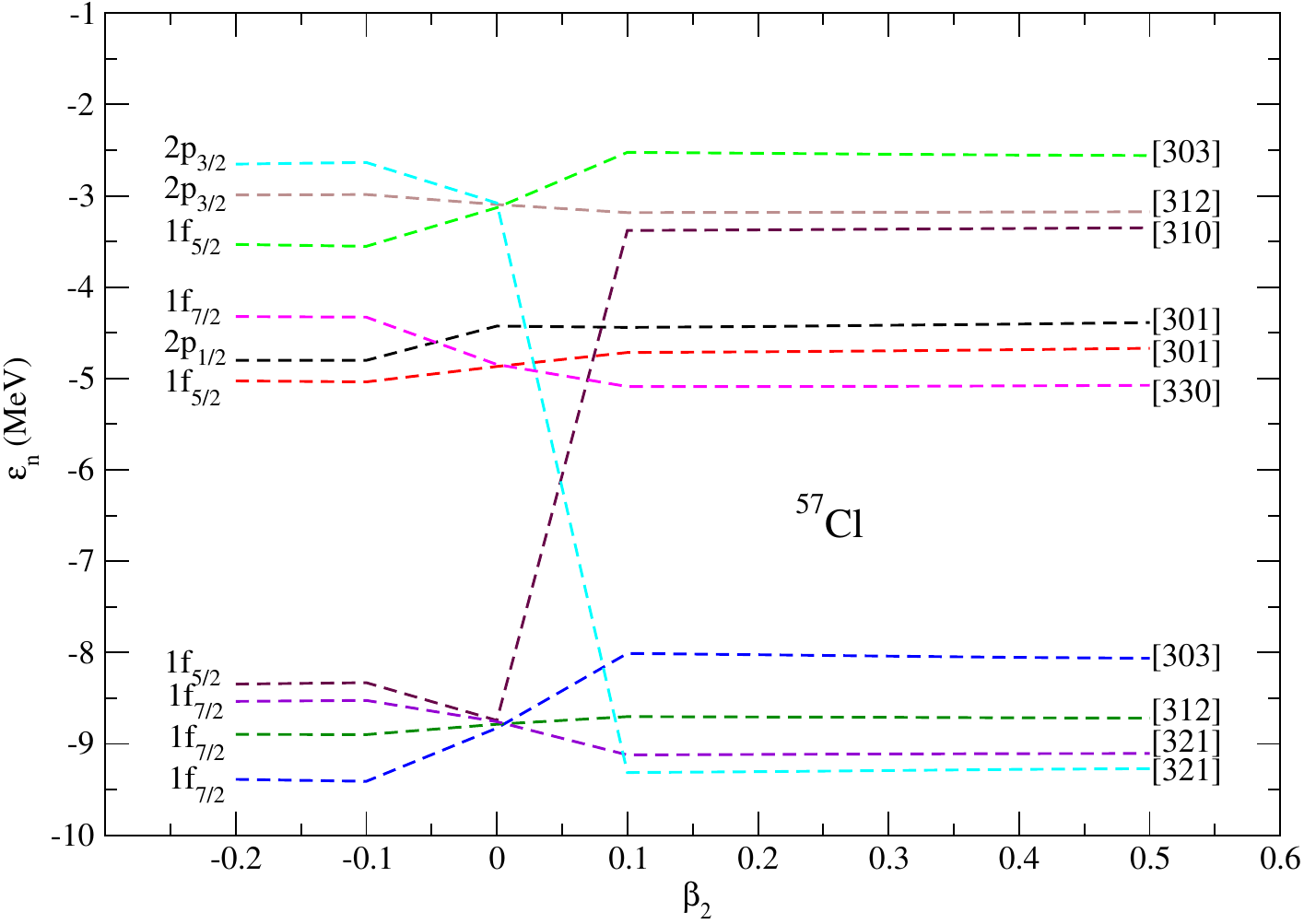}
\caption{The neutron single-particle energies of various nucleonic orbitals of $^{57}$Cl nucleus as a function of quadrupole deformation parameter ($\beta_2$).}
\label{fig8}
\end{figure}
To gain deeper insight into the nature of shell gaps and their dependence on nuclear shape, Fig. \ref{fig8} presents the variation of single-particle energy levels as a function of quadrupole deformation ($\beta_2$) for the nucleus $^{57}$Cl. This figure illustrates how the energies of neutron orbitals evolve as the nucleus transitions from spherical to deformed configurations. The single-particle levels, plotted in MeV, represent individual nucleonic states that respond sensitively to changes in the nuclear shape. As deformation increases, the initially well-separated energy levels begin to shift, with some orbitals approaching one another (level clustering) while others diverge (level repulsion). This deformation-driven evolution of the single-particle spectrum can lead to level crossings and reordering of orbitals, which significantly alter the shell structure compared to the spherical case. In the specific case of $^{57}$Cl, the behaviour of levels near the Fermi surface highlights how deformation can either enhance or weaken an apparent shell gap depending on the arrangement of nearby orbitals. This analysis reveals that the shell gap at N = 34, while visible in the spherical configuration, can be modulated by deformation effects. In some cases, deformation stabilises the nucleus by lowering specific orbitals, favouring certain shapes over others. The energy landscape shown in Fig. 8 also helps explain phenomena such as shape coexistence or sudden changes in deformation across isotopic chains. Thus, studying the evolution of single-particle energy levels with deformation provides a more dynamic picture of nuclear shell structure, especially for neutron-rich and exotic nuclei. It demonstrates that shell gaps are not static features but can be influenced by collective effects, reinforcing the importance of deformation in understanding ground-state configurations and the stability of nuclei far from stability.
\begin{figure}
\centering
\includegraphics[width=1.19\columnwidth,clip=true]{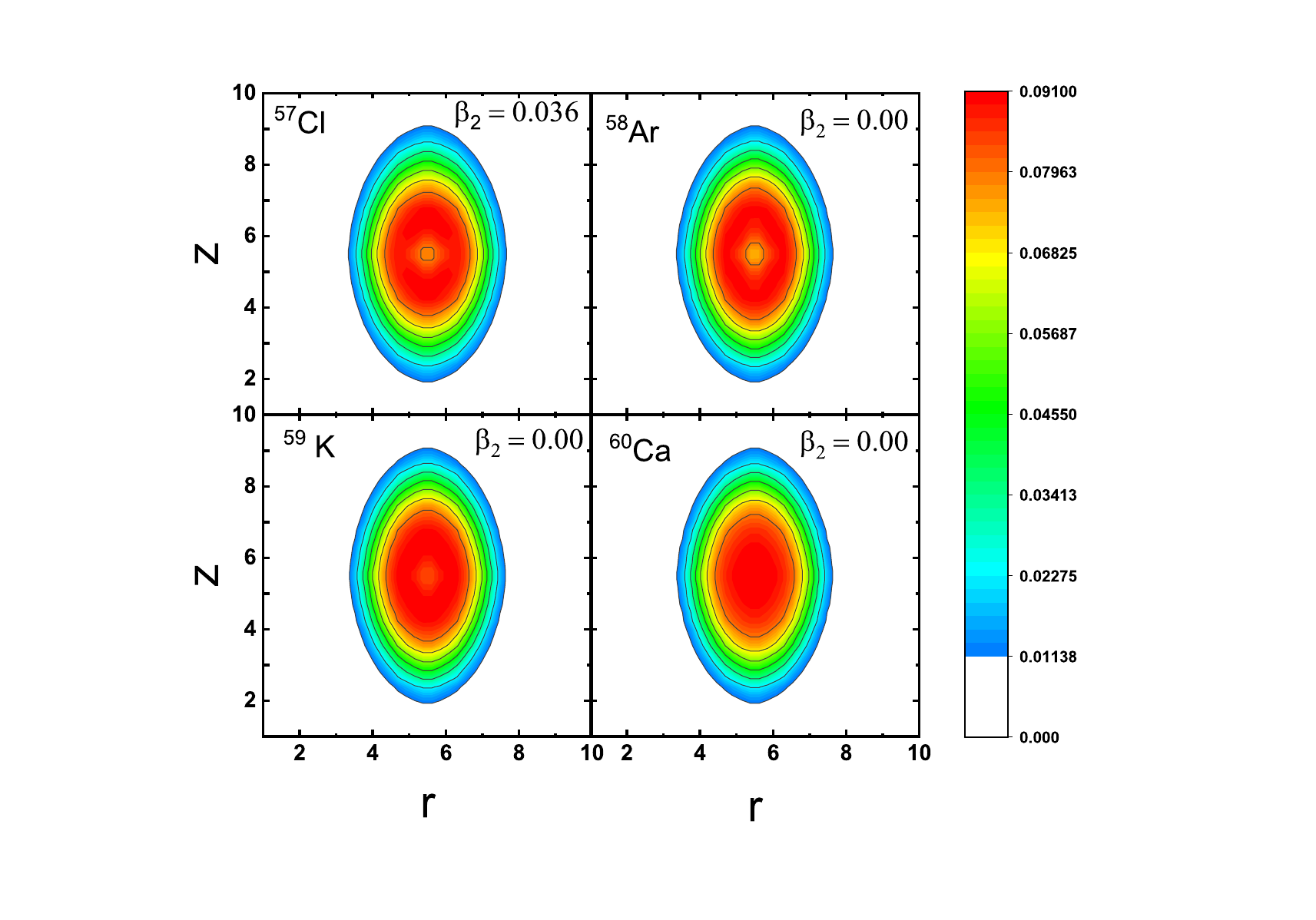}
\caption{The contour plot of the density for the isotopes of Cl, Ar, K and Ca at N = 40.}
\label{fig9}
\end{figure}
\begin{figure}
\centering
\includegraphics[width=1.19\columnwidth,clip=true]{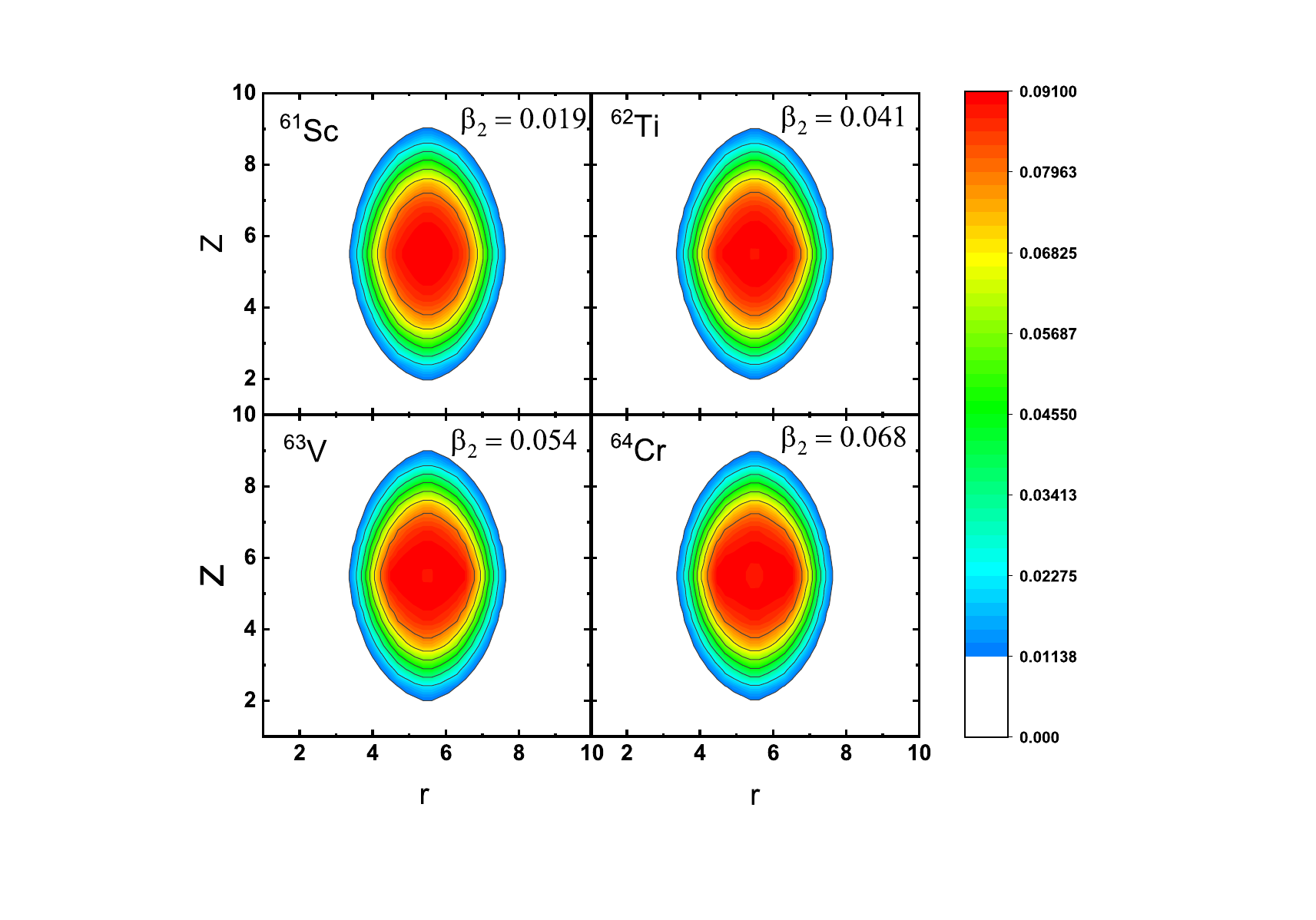}
\caption{Same as \ref{fig9} but for the isotopes of Sc, Ti, V and Cr at N = 40.}
\label{fig10}
\end{figure}

\subsection{3-dimensional density contour}
The systematic progression of density profiles from Cl to Cr nuclei at N=40, as illustrated in Fig. \ref{fig9}, offers valuable insight into the evolving structural characteristics and ground-state configurations across this isotopic chain. These 3-dimensional density contours capture the spatial distribution of nucleons within each nucleus, revealing how nuclear shape, central density, and surface diffuseness change with increasing proton number. This visual representation reflects not just the bulk nuclear properties but also the subtle interplay of shell effects, deformation, and pairing interactions that shape the stability of these neutron-rich nuclei. For lighter elements like Cl and Ar, the density distributions tend to be more centrally peaked and symmetric, indicating nearly spherical or weakly deformed configurations. As we move to heavier nuclei such as Sc, Ti, V, and Cr, the density contours become increasingly elongated or flattened, suggesting a transition to more deformed shapes. These deformations correspond to changes in the underlying shell structure, particularly as the occupation of higher-lying neutron orbitals around N=40 induces collective behaviour and modifies the energy landscape.

Moreover, variations in the density gradients—especially at the nuclear surface—reflect changes in surface thickness and diffuseness, which are closely tied to the symmetry energy and neutron skin development in neutron-rich nuclei. The contour plots thus not only visualize structural evolution but also hint at the changing isospin composition and the effects of isovector interactions. Overall, Fig. \ref{fig9} serves as a powerful diagnostic for studying the ground-state properties of nuclei across the Cl to Cr range. By capturing the complex interdependence of deformation, shell closures, and nucleon distribution, the density profiles provide a foundational basis for understanding the stability and structure of exotic nuclei, particularly near the semi-magic number N=40. These insights contribute significantly to refining nuclear models and interpreting experimental data on shape evolution and shell behaviour far from stability.
\begin{figure}[htbp]
\centering
\includegraphics[width=0.8\columnwidth]{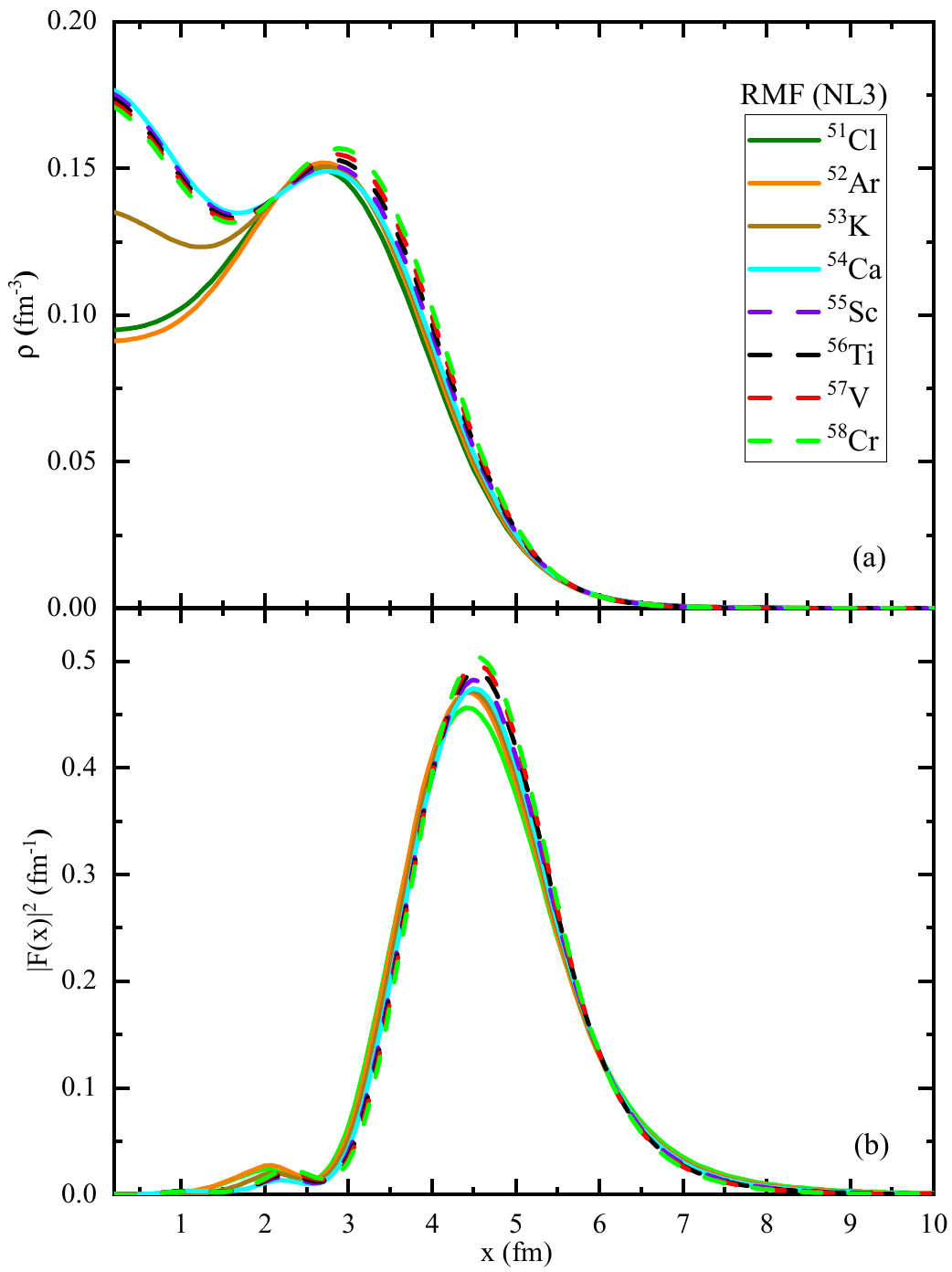}
\caption{\label{fig:dens_fx2}{Total density distribution $\rho$ upper panel (a) and corresponding weight function $|\mathcal{F}(x)|^{2}$ lower panel (b) as a function of nuclear distance using NL3 parameter set for $^{51}$Cl, $^{52}$Ar, $^{53}$K, $^{54}$Ca, $^{55}$Sc, $^{56}$Ti, $^{57}$V and $^{58}$Cr nuclei as representative cases.}}
\end{figure}

\subsection{Density and weight function profile} \label{sec:density_fx2}
The analysis presented in Figs. \ref{fig1}–\ref{fig9} relies on conventional bulk and intrinsic nuclear properties—such as binding energy, charge radius, deformation, and single-particle levels—which are effective in describing nuclei near the valley of $\beta-$stability. However, as we move toward the neutron-rich drip-line, these traditional observables alone may not fully capture the nuances of the nuclear structure due to increasing isospin asymmetry. In such regimes, isospin-dependent quantities become essential for understanding the evolution of shell structure and nuclear stability. To extend our understanding beyond stable nuclei, we incorporate isospin-sensitive observables derived from nucleon density profiles. These profiles offer critical insights into the internal distribution of matter, particularly in neutron-rich systems where the asymmetry between neutron and proton densities becomes significant. The density distribution directly influences various nuclear properties such as the neutron skin thickness, surface symmetry energy, and the behaviour of the symmetry energy, all of which play a crucial role in describing exotic systems.

Fig. \ref{fig:dens_fx2} provides a deeper look into this aspect by presenting RMF (NL3) results for nuclei in the vicinity of the N=28 shell closure for $^{51}$Cl to $^{58}$Cr. Panel (a) illustrates the radial nucleon density profile $\rho (x)$, showing how the density profile varies across this isotopic chain. The gradual extension of the neutron density in more neutron-rich nuclei reflects the development of a neutron-rich surface or neutron skin, a hallmark of increasing isospin asymmetry. Panel (b) of Fig.~\ref{fig:dens_fx2} displays the corresponding weight function $\left(\vert F(x)\vert^2 \right)_x=r$, a key component in the Coherent Density Fluctuation Model (CDFM). This function connects the local density distribution of finite nuclei to infinite nuclear matter properties, allowing for the extraction of symmetry energy and its components. The structure of the weight function highlights the spatial regions contributing most significantly to the nuclear surface, emphasising the importance of these regions in determining isospin-dependent characteristics. Together, the nucleon density profiles and weight functions provide a powerful framework for probing the surface structure of nuclei, particularly in the context of isospin effects. This extended analysis not only complements the traditional bulk property investigation but also enhances our understanding of nuclear matter in neutron-rich environments, especially near magic and semi-magic numbers such as N = 28, 34, and 40.

Panel (a) illustrates the total nucleon density profiles, $\rho(x)$. The distributions exhibit features characteristic of finite nuclei: a relatively saturated density in the nuclear interior followed by a diffuse surface where the density decreases rapidly. For most of the nuclei shown, the central density is approximately $0.15$ fm$^{-3}$. A slight central depression is visible in some profiles, a common feature in mean-field calculations for nuclei in this mass region, arising from shell effects. Despite the variation in proton and neutron numbers across this chain of nuclides, the density profiles are remarkably similar. This indicates that for nuclei with close mass numbers, the overall matter distribution is largely conserved. Minor differences are observable, particularly a subtle outward shift of the nuclear surface for the heavier nuclei (e.g., $^{58}$Cr compared to $^{51}$Cl), reflecting a gentle increase in the nuclear radius with mass number~$A$. Panel (b) provides a complementary perspective by plotting the weight function, $\vert \mathcal{F}(x) \vert^{2}$. This function is typically employed in theoretical formalisms, including the presently used CDFM, to calculate the surface properties namely the symmetry energy and its components by integrating local quantities following Eq. (\ref{eqn:Sf}). The shape and location of $\vert \mathcal{F}(x) \vert^{2}$ reveal which regions of the nucleus contribute most significantly to these properties. The weight function for all considered nuclei is characterized by a pronounced, bell-shaped peak and is negligible at the nuclear center ($x=0$) and in the far exterior.

The crucial insight is gained by comparing the two panels. The peak of the weight function $\vert \mathcal{F}(x) \vert^{2}$ is located at a radial distance of approximately $x \approx 4-5 \, \text{fm}$. Referring back to panel (a), this radial distance corresponds to the nuclear surface, the region where the nucleon density $\rho(x)$ is rapidly falling from its saturation value. Specifically, the peak of $\vert \mathcal{F}(x) \vert^{2}$ occurs where the density is roughly half of the central density, i.e., $\rho(x) \approx 0.07-0.08 \, \text{fm}^{-3}$. This demonstrates that the physical observables calculated using this weight function are predominantly sensitive to the properties of the nuclear surface. The contribution from the high-density interior is strongly suppressed. Therefore, this formalism provides a powerful tool to isolate and study surface phenomena, establishing a direct link between the microscopic density distribution and macroscopic nuclear properties that are governed by the low-density behaviour of nuclear matter.
\begin{figure}[htbp]
\centering
\includegraphics[width=0.8\columnwidth]{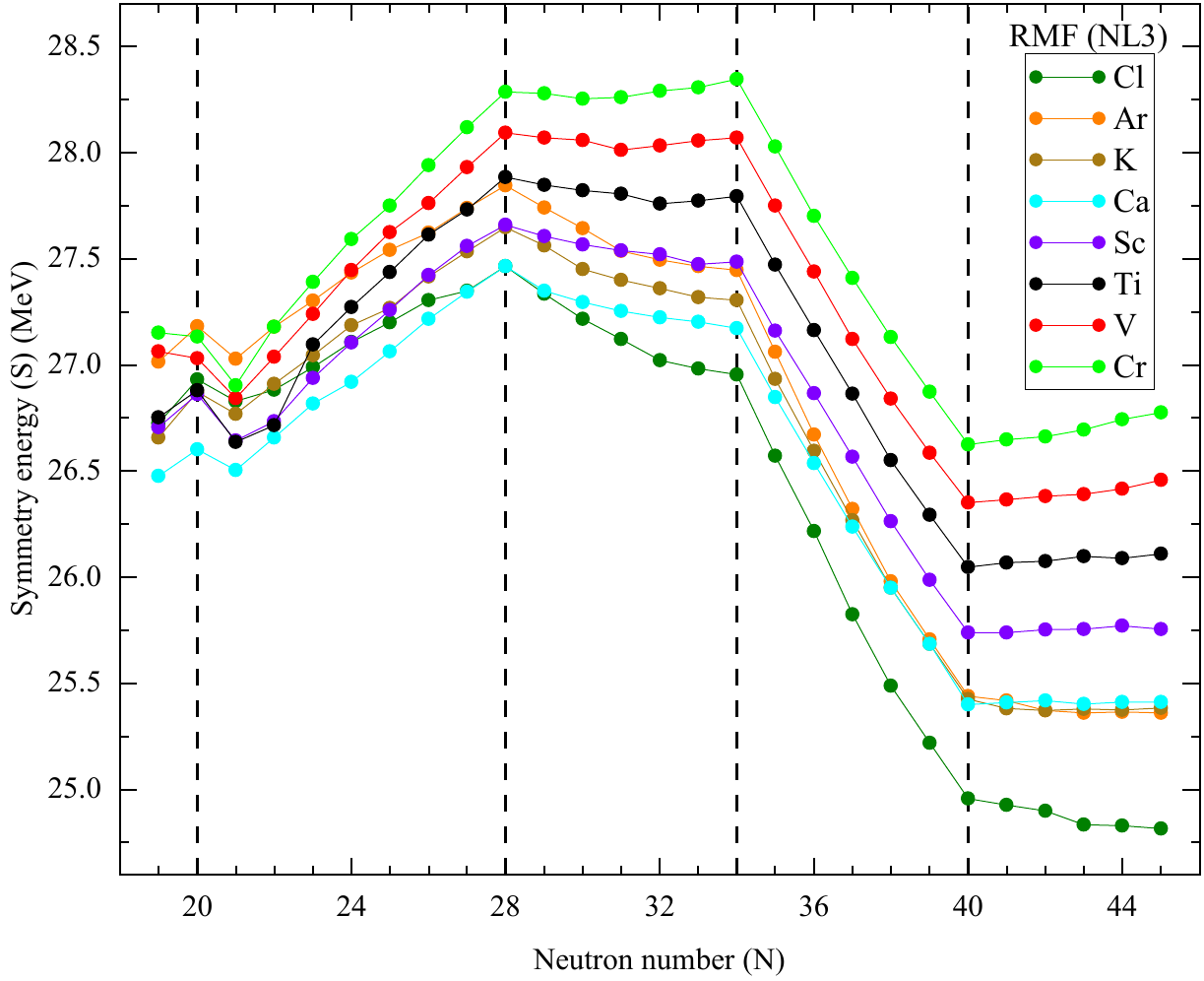}
\caption{\label{fig:symm_energy}{The nuclear symmetry energy of Cl, Ar, K, Ca, Sc, Ti, V and Cr isotopic chains of nuclei using the NL3 parameter set.}}
\end{figure}
\subsection{Nuclear Symmetry Energy}
\label{sec:symmetry_energy}
The nuclear symmetry energy (S) plays a fundamental role in describing the isospin-dependent part of the nuclear equation of state. It quantifies the additional energy required to convert protons into neutrons (or vice versa) in a nucleus, making it a vital observable for understanding the structure of nuclei with large neutron-to-proton asymmetry. This is especially relevant in the study of exotic, neutron-rich systems far from $\beta-$stability, where traditional indicators such as binding energy or separation energy may not fully capture emerging structural features. The symmetry energy is particularly sensitive to changes in nuclear density, shell effects, and the evolution of the neutron skin, making it a powerful tool for probing shell closures and sub-shell effects \cite{bhuy18}. Building on the density profiles and corresponding weight functions 
$\vert \mathcal{F}(x) \vert^{2}$ calculated in Sec. \ref{sec:density_fx2}, we evaluate the symmetry energy $S$ for isotopic chains of Cl, Ar, K, Ca, Sc, Ti, V, and Cr using the Coherent Density Fluctuation Model (CDFM) within the Relativistic Mean-Field (RMF) framework and the NL3 parameterization. The results are depicted in Fig. \ref{fig:symm_energy}, which shows the variation of symmetry energy as a function of neutron number.

These symmetry energy trends reveal distinctive signatures corresponding to known and emerging magic numbers. Notable local maxima or discontinuities are observed at N=20 and N = 28, clearly reflecting the enhanced stability associated with traditional shell closures. Additionally, subtler but consistent enhancements appear at N=34, particularly in the Cl, Ar, and Ti chains, indicating a semi-magic behaviour at this neutron number. This supports earlier findings from bulk observables and single-particle energy level analyses. Crucially, the symmetry energy curves also exhibit a systematic rise near N=40, especially in heavier isotopes such as Ti, V, and Cr. This behaviour reinforces the idea of a shell or sub-shell closure at N=40, consistent with the observed stability in separation energies and the presence of a shell gap in single-particle spectra. The smooth but elevated symmetry energy near N=40 suggests a structural transition in these nuclei, possibly linked to shape coexistence and changes in surface diffuseness due to a thickening neutron skin.

As is evident from Figure~\ref{fig:symm_energy}, the calculations robustly reproduce the well-known magic numbers in this mass region. For all isotopic chains, pronounced peaks or kinks are observed at the neutron numbers $N$ = 20 and $N$ = 28. A local maximum in the symmetry energy indicates that the nucleus is particularly stable at that neutron configuration. The subsequent sharp drop in S for the ($N+1$)th neutron signifies the large energy gap that must be overcome to place a nucleon into the next major shell. This feature confirms the strong shell-quenching effect at these magic numbers. Furthermore, a similar, though less prominent, positive inflection is discernible around $N$ = 34. This feature, visible across multiple isotopic chains, points to the presence of a sub-shell closure. While not as strong as the major shell gaps at $N$ = 20 and 28, this sub-magicity at $N$ = 34 contributes to an island of enhanced stability. This phenomenon is well-documented both theoretically and experimentally \cite{liu20,leis18,leis21,heit24}. The most compelling result from this analysis is the behaviour of the symmetry energy in the vicinity of $N$ = 40. Following the sub-shell closure at $N$ = 34, the symmetry energy for all isotopes exhibits a sharp decrease, as expected when moving further from stability. However, this steep decline is conspicuously arrested as the neutron number approaches 40. Instead of continuing its downward trend, the gradient of the $S$ vs. $N$ curve flattens significantly, culminating in a distinct plateau for $N$ $\geq$ 40.

This plateau is a critical structural indicator. While it does not present as a local maximum like the traditional magic numbers, the abrupt halt in the reduction of symmetry energy is highly significant. It implies that the nucleus resists a further decrease in S, a characteristic feature of an opening energy gap associated with a new shell or sub-shell closure. This behaviour strongly suggests an enhancement in nuclear stability at N = 40 for these neutron-rich systems. The formation of this plateau, interrupting the expected trend, provides compelling evidence for the emergence of N = 40 as a magic number in this region of the nuclear chart. This finding is consistent with observations from other nuclear structure observables, such as two-neutron separation energies discussed in the preceding subsection, which also point towards new magicity at $N$ = 40.
\begin{figure}[htbp]
\centering
\includegraphics[width=0.7\columnwidth]{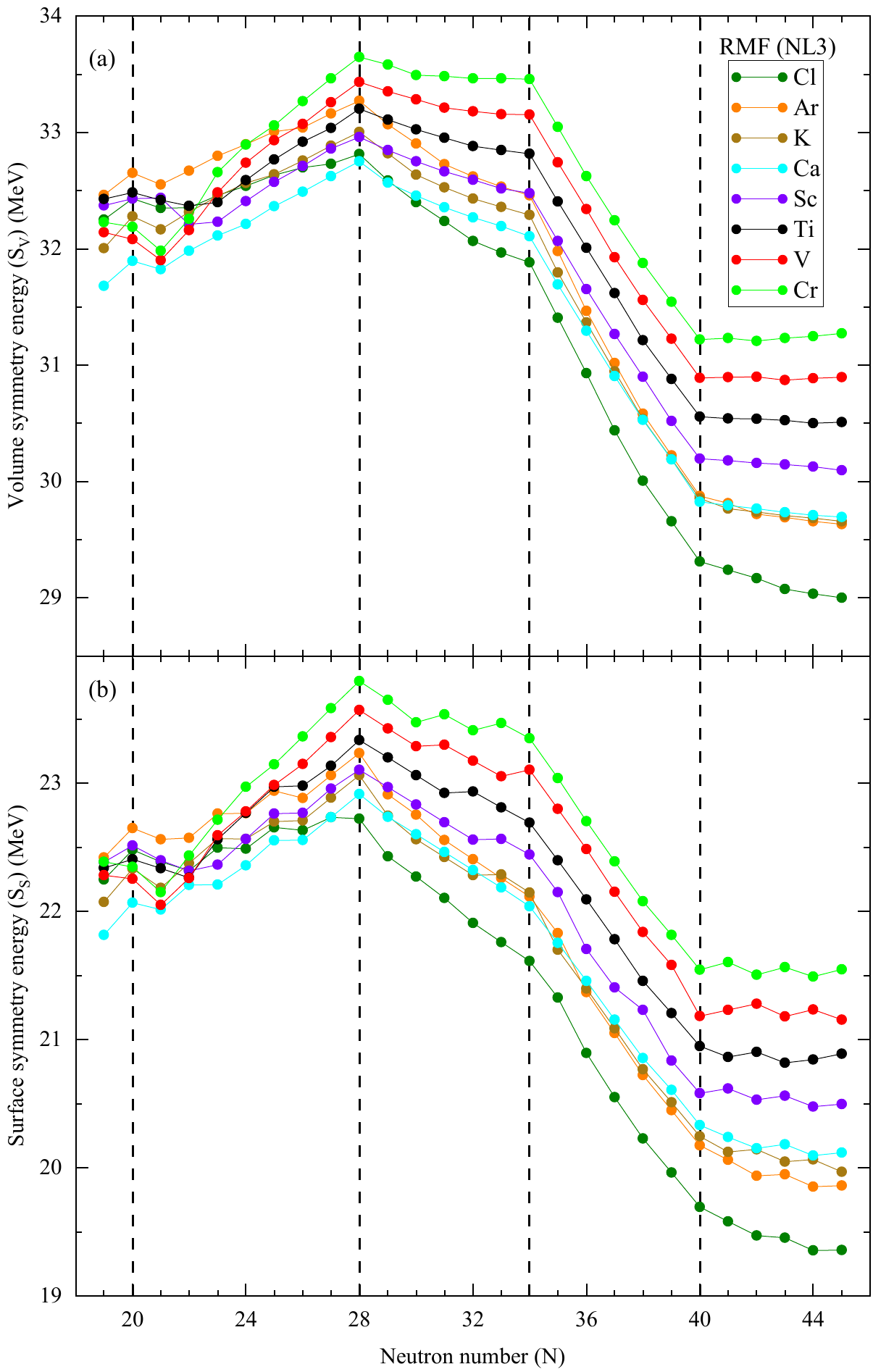}
\caption{\label{fig:sv_ss}{The (a) volume symmetry energy ($S_V$) and (b) surface symmetry energy ($S_S$) as a function of neutron number ($N$) for the isotopic chains of Cl, Ar, K, Ca, Sc, Ti, V, and Cr. Calculations were performed within the RMF framework using the NL3 parameter set.}}
\end{figure}

\subsection{Volume and surface contributions to symmetry energy}
\label{sec:vol_surf_symm_energy}
To gain deeper insight into the origin of the structural effects observed in the nuclear symmetry energy (S), it is instructive to decompose it into its volume ($S_V$) and surface ($S_S$) components as described in Sub-sec. \ref{sec:theory_Sv_Ss}. Following the Danielewicz prescription of liquid-drop \cite{dani06,dani09}, the symmetry energy of a finite nucleus can be expressed as a combination of a bulk term, which is characteristic of infinite nuclear matter, and a surface term, which accounts for the effects of the finite size and diffuse surface of the nucleus. This decomposition allows us to investigate contribution of volume and surface components of symmetry energy in determining the signatures of shell closure. Figure~\ref{fig:sv_ss} presents this decomposition for the Cl-Cr isotopic chains, showing the volume symmetry energy ($S_V$) in panel (a) and the surface symmetry energy ($S_S$) in panel (b), both as a function of neutron number ($N$). A visual inspection immediately reveals that the prominent features observed in the nuclear symmetry energy (Figure~\ref{fig:symm_energy}) are not confined to a single component but are robustly present in both the volume and surface terms.

The well-established shell closures at $N$ = 20 and $N$ = 28 manifest as sharp, pronounced peaks in both the $S_V$ and $S_S$ plots. This indicates that the enhanced stability associated with these magic numbers is a comprehensive effect, influencing the binding energy properties of both the nuclear bulk and its surface region. Similarly, the sub-shell closure at $N$ = 34 is also clearly reflected as a distinct, albeit smaller, peak in both components, reinforcing its characteristic as a genuine structural feature. Most significantly, the emergent shell closure at $N$ = 40 is also clearly delineated in this decomposition. The distinct plateau that interrupts the otherwise decreasing trend in the nuclear symmetry energy is mirrored in both the volume and surface components. As seen in Figure~\ref{fig:sv_ss}(a) and (b), the steep fall-off in both $S_V$ and $S_S$ after $N$ = 34 is arrested around $N$ = 39, leading to the formation of a stable plateau for $N \geq 40$. The fact that this signature of enhanced stability appears independently in both the volume and surface contributions provides powerful, corroborating evidence for the magicity of $N$ = 40 in these neutron-rich nuclei. It demonstrates that the underlying shell-gap formation is a global phenomenon affecting the entire nucleus, rather than being a localized surface or bulk effect.
\section{Conclusion} \label{summary}
In this study, we employ the axially deformed Relativistic Mean Field (RMF) model with the well-established NL3 parameter set to investigate the ground-state properties of isotopic chains ranging from chlorine (Cl) to chromium (Cr). Specifically, we compute key nuclear observables including the charge radius, two-neutron separation energy ($S_{2n}$), quadrupole deformation parameter ($\beta_2$), and single-particle energy levels, which together offer a comprehensive understanding of the shell structure evolution in neutron-rich nuclei. Our analysis reveals clear signatures of shell closures at specific neutron numbers. Most notably, the magic nature of N=40 emerges strongly in the calculated separation energies, where a noticeable drop is observed, indicative of enhanced nuclear stability. This feature is further supported by the presence of a prominent shell gap in the single-particle energy levels, reinforcing the interpretation of N=40 as a robust shell closure in this mass region. In contrast, the behaviour around N = 34 is more nuanced. For lighter isotopes such as $^{51}$Cl, $^{52}$ Ar, $^{53}$K, $^{54}$Ca, and $^{55}$Sc, the two-neutron separation energy does not exhibit a marked drop at N = 34, suggesting weaker or absent shell effects. However, in heavier nuclei such as $^{56}$Ti, $^{57}$V, and $^{58}$Cr, a visible decrease in $S_{2n}$ is evident, which coincides with the development of a shell gap in the corresponding single-particle spectra. This suggests that the manifestation of the N = 34 sub-shell closure becomes more pronounced with increasing proton number, possibly due to the influence of the proton-neutron interaction and the tensor force.

To further substantiate these observations, we analyse the nuclear symmetry energy (S) and its decomposition into volume and surface contributions using the Coherent Density Fluctuation Model (CDFM) based on RMF density distributions. The symmetry energy systematics show distinct features that align well with established and emerging shell closures. Notably, the symmetry energy curve displays local enhancements or plateau-like behaviour at N = 20, 28, 34, and 40, aligning with the points of increased nuclear stability observed in other bulk properties. The decomposition into volume and surface components confirms that these shell effects are not confined to central density regions but are also reflected in the surface structure of the nucleus—a critical aspect for nuclei approaching the drip line. In summary, our RMF+CDFM-based investigation confirms the traditional magic numbers N = 20 and 28, supports the growing body of evidence for sub-magicity at N = 34, and provides strong theoretical backing for the emergence of a new magic number at N = 40. The consistency of these shell features across multiple observables underscores their fundamental nature in the nuclear many-body system and enhances our understanding of structural evolution in neutron-rich nuclei.
\section*{Acknowledgements}
This work is partially supported by the Anusandhan National Research Foundation (ANRF) under the Ramanujan Fellowship, File No. RJF/2022/000140.


\section*{References}\label{bibby}

\begin{thebibliography}{99}
\bibitem{doba94} Dobaczewski J {\it et al} 1994 {\it 
	Phys. Rev. Lett.} {\bf 72}, 981
\bibitem{wern94} Werner T R, Sheikh J A {\it et al} 1994 
		{\it Phys. Lett. B} {\bf 335} 259
\bibitem{chou95} Chou W-T, Casten R F and Zamfir N V 1995 
	{\it Phys. Rev. C} {\bf 51} 2444
\bibitem{ren94} Ren Z, Zhu Y, Cai Y H and Xu G, 
	1994 {\it Phys. Lett. B} {\bf 380} 241
\bibitem{gupt97} Gupta R K, Patra S K and Greiner W 1997 
	{\it Mod. Phys. Lett. A} {\bf 12} 1317
\bibitem{patr97} Patra S K, Gupta R K and Greiner W 1997 
	{\it Int. J. Mod. Phys. E} {\bf 6} 641
\bibitem{meht03} Jha T K, Mehta M S, Patra S K, Raj B K and 
	Gupta R K 2003 {\it Pramana, J. Phys } {\bf 61} 517
\bibitem{mill19} Miller G A, Beck A {\it et al} 2019 
	{\it Phys. Lett. B} {\bf 793} 360
\bibitem{wien13} Wienholtz F, Beck D {\it et al} 2013 
	{\it Nature} {\bf 498} 346
\bibitem{huck85} Huck A, Klotz G {\it et al}1985 
	{\it Phys. Rev. C }{\bf 31} 2226
\bibitem{sorl08} Sorlin O and Porquet M -G 2008 
	{\it Prog. Part. Nucl. Phys.}{\bf 61} 602
\bibitem{ozaw20} Ozawa A, {\it et al} 2000 
	{\it Phys. Rev. Lett.} {\bf 84}	5493
\bibitem{kanu02} Kanungo R, Tanihata I and Ozawa A 2002 
	{\it Phys. Lett. B} {\bf 528} 58
\bibitem{guill84} Guillemaud-Mueller D, {\it et al} 1984 
	{\it Nucl. Phys. A}{\bf 426} 37
\bibitem{moto95} Motobayashi T, {\it et al} 1995 
	{\it Phys. Lett. B}{\bf 346} 9
\bibitem{glas97} Glasmacher T, {\it et al} 1997 
	{\it Phys. Lett. B} {\bf 395} 164
\bibitem{simo99} Simon H, {\it et al} 1999 
	{\it Phys. Rev. Lett.} {\bf  83} 496
\bibitem{navi00} Navin A, {\it et al} 2000 
	{\it Phys. Rev. Lett.} {\bf  85} 266
\bibitem{pris01} Prisciandaro J I, {\it et al} 2001 
	{\it Phys. Lett. B} {\bf 510} 17 
\bibitem{honm05} Honma M, {\it et al} 2005 
	{\it Phys. J. A} {\bf 25 s01}  499
\bibitem{forn04} Fornal B, {\it et al} 2004 
	{\it Phys. Rev. C} {\bf 70} 064304
\bibitem {brod95} Broda R, {\it et al} 1995 
	{\it Phys. Rev. Lett.} {\bf 74} 868
\bibitem{naka08}  Nakada H, 2008 
	{\it Phys. Rev. C} {\bf 78} 054301
\bibitem{naka10} Nakada H, 2010 
	{\it Phys. Rev. C} {\bf 81} 027301
\bibitem{tera06} Terasaki J and Engel J, 2006 
	{\it Phys. Rev. C} {\bf 74} 044301
\bibitem{iimu23} Iimura S, {\it et al} 2023 
	{\it Phys. Rev. Lett.}{\bf  130} 012501
\bibitem{malb22} Malbrunot-Ettenauer S, {\it et al} 2022 
	{\it Phys. Rev. Lett.} {\bf 128} 022502
\bibitem{groo20} Groote R P de, {\it et al} 2020 
	{\it Nature Phys.} {\bf 16} 620
\bibitem{babc16} Babcock C {\it et al} 2016 
	{\it Phys. Lett. B} {\bf 760} 387
\bibitem{biss16} Bissell M L, {\it et al} 2016 
	{\it Phys. Rev. C} {\bf 93} 064318
\bibitem{moug18} Mougeot M, {\it et al} 2018 
	{\it Phys. Rev. Lett.} {\bf 120} 232501
\bibitem{tani19} Taniuchi R, {\it et al} 2019 
	{\it Nature Phys.} {\bf 569} 53
\bibitem{nowa16} Nowacki F, {\it et al} 2016 
	{\it  Phys. Rev. Lett.} {\bf  117} 272501
\bibitem{somm22} Sommer F, {\it et al} 2022 
	{\it Phys. Rev. Lett.} {\bf 129} 132501
\bibitem{satp04} Satpathy L and Patra S K, 2004 
	{\it J. Phys. G: Nucl. Part. Phys.} {\bf 30} 771
\bibitem{gaid11} Gaidarov M, Antonov A, Sarriguren P and 
	Guerra E M de, 2011 {\it Phys. Rev. C} \textbf{84} 
	034316
\bibitem{gaid12} Gaidarov M, Antonov A, Sarriguren P and 
	Guerra E M de, 2012 {\it Phys. Rev. C} \textbf{85} 
	064319
\bibitem{rufa88} Rufa M, Reinhard P -G, Maruhn J A, Greiner W 
	and Strayer M R, 1988 {\it Phys. Rev. C} {\bf 38} 390
\bibitem{hard88} Reinhard P -G, 1988 
	{\it Z. Phys. A} {\bf 329} 257
\bibitem{bhuy18} Bhuyan M, Carlson B, Patra S K and Zhou S -G, 
	2028 {\it Phys. Rev. C} \textbf{97} 024322
\bibitem{patr09} Patra S K, Bhuyan M, Mehta M S and Gupta R K, 
	2009 {\it Phys. Rev. C} {\bf 80} 034312
\bibitem{type01} Typel S and  Brown B A, 2001 
	{\it Phys. Rev. C} {\bf 64} 027302 
\bibitem{rein86} Reinhard P -G, Rufa M, Maruhn J A, Greiner W 
	and Friedrich J, 1986 {\it Z. Phys. A} {\bf 323} 13
\bibitem{bhuy15} Bhuyan M, 2015 {\it Phys. Rev. C} {\bf 92} 
	034323 
\bibitem{anto16} Antonov A, Gaidarov M, Sarriguren P and 
	Guerra E M de, 2016 {\it Phys. Rev. C} \textbf{94} 
	014319
\bibitem{prav22} Yadav P K, Kumar R and Bhuyan M, 2022 
	{\it Chin. Phys. C} \textbf{46} 084101
\bibitem{prav_25npa} Biswal N, Yadav P K, Panda R N, Mishra S 
	and Bhuyan M, 2025 {\it Nucl. Phys. A} \textbf{1053} 
	122975
\bibitem{myer80} Myers W D and Swiatecki M J, 1980 
	{\it Nucl. Phys. A} \textbf{336} 267
\bibitem{dani06} Danielewicz P, 2007 Opportunities with 
	Exotic Beams (World Scientific) 142
\bibitem{wale74} Walecka J D, 1974 
	{\it Ann. Phys.} {\bf 83} 491
\bibitem{sero86} Serot B D and  Walecka J D, 1986 
	{\it The Relativistic Nuclear Many Body Problem}
	{\bf 16}
\bibitem{bhuy12} Singh B B, Bhuyan M, Patra S K and GuptaR K,
        2012 {\it J. Phys. G: Nucl. and Part. Phys.} {\bf 39(2)} 025101
\bibitem{madl88} Madland D G and Nix J R, 1988 
	{\it Nucl. Phys.A } {\bf 476} 1
\bibitem{moll88} Moller P and Nix J R, 1988 
	{\it At. Data and Nucl. Data Tables} {\bf 39} 213
\bibitem{patr93} Patra S K, 1993 
	{\it Phys. Rev. C} {\bf 48} 1449 
\bibitem{pres82} Preston M A and Bhaduri R K, 1982 
	{\it Structure of Nucleus (Addison-Wesley, Boston)} 
	Chap. 8 {\bf  309}
\bibitem{gaid21} Gaidarov M, Guerra E M de, Antonov A, 
	Danchev I, Sarriguren P and Kadrev D, 2012 
	{\it Phys. Rev. C} \textbf{104} 044312
\bibitem{anto04} Antonov A, Gaidarov M, Kadrev D, Ivanov M, 
	Guerra E M de and Udias J, 2004 {\it Phys. Rev. C} 
	\textbf{69} 044321
\bibitem{anto05} Antonov A, Gaidarov M, Ivanov M, Kadrev D, 
	Guerra E M de, Sarriguren P and Udias J, 2005 
	{\it Phys. Rev. C} \textbf{71} 014317
\bibitem{anto07} Antonov A, Ivanov M, Barbaro M B, 
	Caballero J, Guerra E M de and Gaidarov M, 2007 
	{\it Phys. Rev. C} \textbf{75} 064617
\bibitem{ivan08} Ivanov M, Barbaro M B, Caballero J, Antonov 
	A, Guerra E M de and Gaidarov M, 2008 
	{\it Phys. Rev. C} \textbf{77} 034612
\bibitem{prav24} Yadav P K, Kumar R and Bhuyan M, 2024 
	{\it Europhys. Lett.} \textbf{146} 14001
\bibitem{prav23} Yadav P K, Kumar R and Bhuyan M, 2023 
	{\it Mod. Phys. Lett. A} \textbf{38} 2350114
\bibitem{gamb90} Gambhir Y K, Ring P and  Thimet A, 1990 
	{\it Ann. Phys.} (N.Y.) {\bf 198} 132
\bibitem{horo81} Horowitz C J and Serot B D, 1981 
	{\it Nucl. Phys. A} {\bf 368} 503
\bibitem{pann87} Pannert W, Ring P and Boguta J, 1987 
	{\it Phy. Rev. Lett.} {\bf 59}, 2420
\bibitem{doba84} Dobaczewski J, Flocard H and Treiner J, 1984 
	{\it Nucl. Phys. A} {\bf 422} 103
\bibitem{patr01} Patra S K, Estal M Del, Centelles M and 
	Vinas X, 2001 {\it Phys. Rev.C} {\bf 63} 024311
\bibitem{meht02} Mehta M S, Raj B K, Patra S K and Gupta R K, 2002
	{\it Phys. Rev.C} {\bf 66} 044317
\bibitem{saho20} Sahoo T and Patra S K, 2020 
	{\it Phys. Scr.} {\bf 95} 085302
\bibitem{wang21} Wang M, {\it et al} 2021 
	{\it Chinese Physics C} {\bf 45} 030003
\bibitem{prit16} Pritychenko B, {\it et al} 2016 
	{\it Atomic Data and Nuclear Data Tables} {\bf 107} 1
\bibitem{adri20} Adri M El and Oulne M, 2020 
	{\it Th European Phys. J. Plus} {\bf 135} 268
\bibitem{ange13} Angeli I, Marinova A, 2013 At. Data and Nucl. Data Table {\bf 99} 69
\bibitem{anto80} Antonov A, Nikolaev V, and Petkov I Z, 
	Zeitschrift, 1980 
	f{\"u}r Physik A Atoms and Nuclei \textbf{297} 257
\bibitem{anto79} Antonov A, Nikolaev V and Petkov I Z, 1979 
	Bulg. J. Phys. \textbf{6}
\bibitem{anto82} Antonov A, Nikolaev V and Petkov I Z, 1982 
	Zeitschrift f{\"u}r Physik A Atoms and Nuclei 
	\textbf{304} 239
\bibitem{grif57} Griffin J J and Wheeler J A, 1957 
	{\it Phys. Rev.} \textbf{108} 311
\bibitem{anto94} Antonov A, Kadrev D and Hodgson P, 1994
	{\it Phys. Rev. C} \textbf{50} 164
\bibitem{hohe64} Hohenberg P and Kohn W, 1964 
	{\it Phys. Rev.} \textbf{136} B864
\bibitem{brue68} Brueckner K, Buchler J, Jorna S and Lombard 
	R, 1968 {\it Phys. Rev.} \textbf{171} 1188
\bibitem{brue69} Brueckner K, Buchler J, Clark R and Lombard 
	R, 1969 {\it Phys. Rev.} \textbf{181} 1543
\bibitem{sarr07} Sarriguren P, Gaidarov M, Guerra E M de 
	and Antonov A, 2007 {\it Phys. Rev. C} \textbf{76} 
	044322
\bibitem{dani03} Danielewicz P, 2003 {\it Nucl. Phys. A} 
	\textbf{727} 233 
\bibitem{dani09} Danielewicz P and Lee J, 2009 {\it Nucl. 
	Phys. A} \textbf{818} 36
\bibitem{danc20} Danchev I, Antonov A, Kadrev D, Gaidarov M, 
	Sarriguren P and Guerra E M de, 2020 
	{\it Phys. Rev. C} \textbf{101} 064315
\bibitem{mo15} Mo Q, Liu M, Cheng L and Wang N, 2015 
	{\it Sci. Chin. Phys., Mech. \& Astro.} 
	\textbf{58} 1 
\bibitem{diep07} Dieperink A E L and Isacker P Van, 2007 
	{\it Eur. Phys. J. A} {\bf 32} 11

\bibitem{liu20} Liu J, Niu Y F, Long W H, 2020 
	{\it Phys. Lett. B} \textbf{806} 135524
\bibitem{leis18} Leistenschneider E {\it et al}, 2018 
	{\it Physical Rev. Lett.} \textbf{120} 062503 
\bibitem{leis21} Leistenschneider E {\it et al}, 2021 
	{\it Physical Rev. Lett.} \textbf{126} 042501
\bibitem{heit24} Heitz L, EbranJ -P, Khan E and Verney D, 2024 
	{\it arXiv:2411.15562 [nucl-th]}


\end{thebibliography}
\end{document}